\documentclass[a4paper,11pt]{article}
\usepackage{jcappub}
\usepackage{fontawesome}
\usepackage{orcidlink}
\usepackage[modulo]{lineno}

\newcommand{\lya}{Ly\ensuremath{\alpha}}

\newcommand{\kms}{\text{s$\cdot$km}^{-1}}
\newcommand{\invkms}{\text{km$\cdot$s}^{-1}}
\newcommand{\invAA}{\AA$^{-1}$}
\newcommand{\hi}{H{\textsc{i}}}
\newcommand{\siii}{Si{\textsc{ii}}}
\newcommand{\siiii}{Si{\textsc{iii}}}
\newcommand{\siiv}{Si{\textsc{iv}}}
\newcommand{\mgii}{Mg{\textsc{ii}}}
\newcommand{\civ}{C{\textsc{iv}}}
\newcommand{\pk}{P_{1\mathrm{D},\alpha}}
\newcommand{\pkcross}{P_{1\mathrm{D},\alpha,\mathrm{cross}}}
\newcommand{\pns}{P_{\mathrm{noise},s}}
\newcommand{\prs}{P_{\mathrm{raw},s}}
\newcommand{\ps}{P_{s}}
\newcommand{\pn}{\left\langle \pns(k) \right\rangle_{s \in z, k\in A}}
\newcommand{\pr}{\left\langle \prs(k) \right\rangle_{s \in z, k\in A}}
\newcommand{\pme}{P_{\mathrm{metals}}}
\newcommand{\psbm}{P_{\mathrm{SB1,m}}}
\newcommand{\psbmz}{P_{\mathrm{SB1,m,z}}}
\newcommand{\snr}{\overline{\mathrm{SNR}}}
\newcommand{\cnr}{\overline{\mathrm{CNR}}}

\newcommand{\rmats}{R_s}
\newcommand{\dD}{\delta_{\mathrm{D}}}
\newcommand{\df}{\delta_{F}}
\newcommand{\dfs}{\delta_{F,s}}
\newcommand{\dlya}{\delta_{\mathrm{\lya}}}
\newcommand{\dlyas}{\delta_{\mathrm{\lya},s}}
\newcommand{\dnoises}{\delta_{\mathrm{noise},s}}
\newcommand{\dsiii}{\delta_{\mathrm{\siii}}}
\newcommand{\dsiiii}{\delta_{\mathrm{\siiii}}}
\newcommand{\done}{\delta_1}
\newcommand{\dtwo}{\delta_2}
\newcommand{\vecdone}{\vec{\delta}_1}
\newcommand{\vecdtwo}{\vec{\delta}_2}
\newcommand{\vecdfs}{\vec{\delta}_{F,s}}
\newcommand{\vecdlyas}{\vec{\delta}_{\mathrm{\lya},s}}
\newcommand{\vecdnoises}{\vec{\delta}_{\mathrm{noise},s}}
\newcommand{\vecrmats}{\vec{R}_s}
\newcommand{\vecdme}{\vec{\delta}_{\mathrm{metal}}}
\newcommand{\var}{{\rm var}}
\newcommand{\cov}{{\rm cov}}
\newcommand{\git}[2]{\href{https://github.com/#1}{\faGithub}\footnote{\label{#1}#2\url{https://github.com/#1}}}
\newcommand{\edrmtwo}{EDR+M2}
\newcommand{\drone}{DR1}
\newcommand{\dronefull}{DESI-DR1}
\newcommand{\desiedrfft}{R23}
\newcommand{\validpaper}{KR25}
\newcommand{\qmleyone}{K25}

\arxivnumber{2505.09493}

\title{\boldmath DESI DR1 Ly$\alpha$ 1D power spectrum: The Fast Fourier Transform estimator measurement}

\author[1]{Corentin Ravoux\orcidlink{0000-0002-3500-6635}}
\author[2]{Marie-Lynn Abdul-Karim\orcidlink{0009-0000-7133-142X}}
\author[2]{Jean-Marc Le Goff}
\author[2]{Eric Armengaud\orcidlink{0000-0001-7600-5148}}
\author[3]{{Jessica N. Aguilar}}
\author[4]{{Steven Ahlen}\orcidlink{0000-0001-6098-7247}}
\author[3]{{Stephen Bailey}\orcidlink{0000-0003-4162-6619}}
\author[5,6]{{Davide Bianchi}\orcidlink{0000-0001-9712-0006}}
\author[3]{{Allyson Brodzeller}\orcidlink{0000-0002-8934-0954}}
\author[7]{{David Brooks}}
\author[8]{{Jon\'{a}s Chaves-Montero}\orcidlink{0000-0002-9553-4261}}
\author[3]{{Todd Claybaugh}}
\author[3,9]{{Andrei Cuceu}\orcidlink{0000-0002-2169-0595}}
\author[3]{{Roger de Belsunce}\orcidlink{0000-0003-3660-4028}}
\author[10]{{Axel de la Macorra}\orcidlink{0000-0002-1769-1640}}
\author[11]{{Arjun Dey}\orcidlink{0000-0002-4928-4003}}
\author[12]{{Zhejie Ding}\orcidlink{0000-0002-3369-3718}}
\author[7]{{Peter Doel}}
\author[3,13]{{Simone Ferraro}\orcidlink{0000-0003-4992-7854}}
\author[8]{{Andreu Font-Ribera}\orcidlink{0000-0002-3033-7312}}
\author[14,15]{{Jaime E. Forero-Romero}\orcidlink{0000-0002-2890-3725}}
\author[16,17,18]{{Enrique Gaztañaga}}
\author[19,20,21]{{Naim G\" oksel Kara\c{c}ayl{\i}}\orcidlink{0000-0001-7336-8912}}
\author[3]{{Satya Gontcho A Gontcho}\orcidlink{0000-0003-3142-233X}}
\author[22]{{Gaston Gutierrez}}
\author[3]{{Julien Guy}\orcidlink{0000-0001-9822-6793}}
\author[2,23]{{Hiram K. Herrera-Alcantar}\orcidlink{0000-0002-9136-9609}}
\author[24]{{Mustapha Ishak}\orcidlink{0000-0002-6024-466X}}
\author[25]{{Robert Kehoe}}
\author[26]{{David Kirkby}\orcidlink{0000-0002-8828-5463}}
\author[3]{{Theodore Kisner}\orcidlink{0000-0003-3510-7134}}
\author[3]{{Anthony Kremin}\orcidlink{0000-0001-6356-7424}}
\author[3]{{Martin Landriau}\orcidlink{0000-0003-1838-8528}}
\author[27]{{Laurent Le Guillou}\orcidlink{0000-0001-7178-8868}}
\author[3]{{Michael E. Levi}\orcidlink{0000-0003-1887-1018}}
\author[28,29]{{Marc Manera}\orcidlink{0000-0003-4962-8934}}
\author[19,20]{{Paul Martini}\orcidlink{0000-0002-4279-4182}}
\author[11]{{Aaron Meisner}\orcidlink{0000-0002-1125-7384}}
\author[8,29]{{Ramon Miquel}}
\author[30]{{Paulo Montero-Camacho}\orcidlink{0000-0002-6998-6678}}
\author[10]{{Andrea Muñoz-Gutiérrez}}
\author[17]{{Seshadri Nadathur}\orcidlink{0000-0001-9070-3102}}
\author[31,32]{{Gustavo Niz}\orcidlink{0000-0002-1544-8946}}
\author[2,3]{{Nathalie Palanque-Delabrouille}\orcidlink{0000-0003-3188-784X}}
\author[33]{{Zhiwei Pan}\orcidlink{0000-0003-0230-6436}}
\author[34,35,36]{{Will J. Percival}\orcidlink{0000-0002-0644-5727}}
\author[37]{{Ignasi P\'erez-R\`afols}\orcidlink{0000-0001-6979-0125}}
\author[38]{{Matthew M. Pieri}\orcidlink{0000-0003-0247-8991}}
\author[39]{{Francisco Prada}\orcidlink{0000-0001-7145-8674}}
\author[40]{{Graziano Rossi}}
\author[41]{{Eusebio Sanchez}\orcidlink{0000-0002-9646-8198}}
\author[42]{{Christoph Saulder}\orcidlink{0000-0002-0408-5633}}
\author[3]{{David Schlegel}}
\author[43,44]{{Michael Schubnell}}
\author[45]{{Hee-Jong Seo}\orcidlink{0000-0002-6588-3508}}
\author[3]{{Joseph H. Silber}\orcidlink{0000-0002-3461-0320}}
\author[18,46]{{Małgorzata Siudek}\orcidlink{0000-0002-2949-2155}}
\author[11]{{David Sprayberry}}
\author[2]{{Ting Tan}\orcidlink{0000-0001-8289-1481}}
\author[47]{{Ji-Jia Tang}\orcidlink{0000-0002-1860-0886}}
\author[44]{{Gregory Tarl\'{e}}\orcidlink{0000-0003-1704-0781}}
\author[48,49]{{Michael Walther}\orcidlink{0000-0002-1748-3745}}
\author[11]{{Benjamin A. Weaver}}
\author[2]{{Christophe Yèche}\orcidlink{0000-0001-5146-8533}}
\author[50]{{Jiaxi Yu}\orcidlink{0009-0001-7217-8006}}
\author[3]{{Rongpu Zhou}\orcidlink{0000-0001-5381-4372}}
\author[51]{{Hu Zou}\orcidlink{0000-0002-6684-3997}}

\affiliation[1]{Université Clermont-Auvergne, CNRS, LPCA, 63000 Clermont-Ferrand, France}
\affiliation[2]{IRFU, CEA, Universit{\'e} Paris-Saclay, D36, Gif-sur-Yvette F-91191, France}
\affiliation[3]{Lawrence Berkeley National Laboratory, 1 Cyclotron Road, Berkeley, CA 94720, USA}
\affiliation[4]{Department of Physics, Boston University, 590 Commonwealth Avenue, Boston, MA 02215 USA}
\affiliation[5]{Dipartimento di Fisica ``Aldo Pontremoli'', Universit\`a degli Studi di Milano, Via Celoria 16, I-20133 Milano, Italy}
\affiliation[6]{INAF-Osservatorio Astronomico di Brera, Via Brera 28, 20122 Milano, Italy}
\affiliation[7]{Department of Physics \& Astronomy, University College London, Gower Street, London, WC1E 6BT, UK}
\affiliation[8]{Institut de F\'{i}sica d’Altes Energies (IFAE), The Barcelona Institute of Science and Technology, Edifici Cn, Campus UAB, 08193, Bellaterra (Barcelona), Spain}
\affiliation[9]{NASA Einstein Fellow}
\affiliation[10]{Instituto de F\'{\i}sica, Universidad Nacional Aut\'{o}noma de M\'{e}xico,  Circuito de la Investigaci\'{o}n Cient\'{\i}fica, Ciudad Universitaria, Cd. de M\'{e}xico  C.~P.~04510,  M\'{e}xico}
\affiliation[11]{NSF NOIRLab, 950 N. Cherry Ave., Tucson, AZ 85719, USA}
\affiliation[12]{University of Chinese Academy of Sciences, Nanjing 211135, People's Republic of China.}
\affiliation[13]{University of California, Berkeley, 110 Sproul Hall \#5800 Berkeley, CA 94720, USA}
\affiliation[14]{Departamento de F\'isica, Universidad de los Andes, Cra. 1 No. 18A-10, Edificio Ip, CP 111711, Bogot\'a, Colombia}
\affiliation[15]{Observatorio Astron\'omico, Universidad de los Andes, Cra. 1 No. 18A-10, Edificio H, CP 111711 Bogot\'a, Colombia}
\affiliation[16]{Institut d'Estudis Espacials de Catalunya (IEEC), c/ Esteve Terradas 1, Edifici RDIT, Campus PMT-UPC, 08860 Castelldefels, Spain}
\affiliation[17]{Institute of Cosmology and Gravitation, University of Portsmouth, Dennis Sciama Building, Portsmouth, PO1 3FX, UK}
\affiliation[18]{Institute of Space Sciences, ICE-CSIC, Campus UAB, Carrer de Can Magrans s/n, 08913 Bellaterra, Barcelona, Spain}
\affiliation[19]{Center for Cosmology and AstroParticle Physics, The Ohio State University, 191 West Woodruff Avenue, Columbus, OH 43210, USA}
\affiliation[20]{Department of Astronomy, The Ohio State University, 4055 McPherson Laboratory, 140 W 18th Avenue, Columbus, OH 43210, USA}
\affiliation[21]{Department of Physics, The Ohio State University, 191 West Woodruff Avenue, Columbus, OH 43210, USA}
\affiliation[22]{Fermi National Accelerator Laboratory, PO Box 500, Batavia, IL 60510, USA}
\affiliation[23]{Institut d'Astrophysique de Paris. 98 bis boulevard Arago. 75014 Paris, France}
\affiliation[24]{Department of Physics, The University of Texas at Dallas, 800 W. Campbell Rd., Richardson, TX 75080, USA}
\affiliation[25]{Department of Physics, Southern Methodist University, 3215 Daniel Avenue, Dallas, TX 75275, USA}
\affiliation[26]{Department of Physics and Astronomy, University of California, Irvine, 92697, USA}
\affiliation[27]{Sorbonne Universit\'{e}, CNRS/IN2P3, Laboratoire de Physique Nucl\'{e}aire et de Hautes Energies (LPNHE), FR-75005 Paris, France}
\affiliation[28]{Departament de F\'{i}sica, Serra H\'{u}nter, Universitat Aut\`{o}noma de Barcelona, 08193 Bellaterra (Barcelona), Spain}
\affiliation[29]{Instituci\'{o} Catalana de Recerca i Estudis Avan\c{c}ats, Passeig de Llu\'{\i}s Companys, 23, 08010 Barcelona, Spain}
\affiliation[30]{Department of Mathematics and Theory, Peng Cheng Laboratory, Shenzhen, Guangdong 518066, China}
\affiliation[31]{Departamento de F\'{\i}sica, DCI-Campus Le\'{o}n, Universidad de Guanajuato, Loma del Bosque 103, Le\'{o}n, Guanajuato C.~P.~37150, M\'{e}xico}
\affiliation[32]{Instituto Avanzado de Cosmolog\'{\i}a A.~C., San Marcos 11 - Atenas 202. Magdalena Contreras. Ciudad de M\'{e}xico C.~P.~10720, M\'{e}xico}
\affiliation[33]{Kavli Institute for Astronomy and Astrophysics at Peking University, PKU, 5 Yiheyuan Road, Haidian District, Beijing 100871, P.R. China}
\affiliation[34]{Department of Physics and Astronomy, University of Waterloo, 200 University Ave W, Waterloo, ON N2L 3G1, Canada}
\affiliation[35]{Perimeter Institute for Theoretical Physics, 31 Caroline St. North, Waterloo, ON N2L 2Y5, Canada}
\affiliation[36]{Waterloo Centre for Astrophysics, University of Waterloo, 200 University Ave W, Waterloo, ON N2L 3G1, Canada}
\affiliation[37]{Departament de F\'isica, EEBE, Universitat Polit\`ecnica de Catalunya, c/Eduard Maristany 10, 08930 Barcelona, Spain}
\affiliation[38]{Aix Marseille Univ, CNRS, CNES, LAM, Marseille, France}
\affiliation[39]{Instituto de Astrof\'{i}sica de Andaluc\'{i}a (CSIC), Glorieta de la Astronom\'{i}a, s/n, E-18008 Granada, Spain}
\affiliation[40]{Department of Physics and Astronomy, Sejong University, 209 Neungdong-ro, Gwangjin-gu, Seoul 05006, Republic of Korea}
\affiliation[41]{CIEMAT, Avenida Complutense 40, E-28040 Madrid, Spain}
\affiliation[42]{Max Planck Institute for Extraterrestrial Physics, Gie\ss enbachstra\ss e 1, 85748 Garching, Germany}
\affiliation[43]{Department of Physics, University of Michigan, 450 Church Street, Ann Arbor, MI 48109, USA}
\affiliation[44]{University of Michigan, 450 Church Street, Ann Arbor, MI 48109, USA}
\affiliation[45]{Department of Physics \& Astronomy, Ohio University, 139 University Terrace, Athens, OH 45701, USA}
\affiliation[46]{Instituto de Astrof\'{\i}sica de Canarias, C/ V\'{\i}a L\'{a}ctea, s/n, E-38205 La Laguna, Tenerife, Spain}
\affiliation[47]{Graduate Institute of Astrophysics and Department of Physics, National Taiwan University, No. 1, Sec. 4, Roosevelt Rd., Taipei 10617, Taiwan}
\affiliation[48]{Excellence Cluster ORIGINS, Boltzmannstrasse 2, D-85748 Garching, Germany}
\affiliation[49]{University Observatory, Faculty of Physics, Ludwig-Maximilians-Universit\"{a}t, Scheinerstr. 1, 81677 M\"{u}nchen, Germany}
\affiliation[50]{Institute of Physics, Laboratory of Astrophysics, \'{E}cole Polytechnique F\'{e}d\'{e}rale de Lausanne (EPFL), Observatoire de Sauverny, Chemin Pegasi 51, CH-1290 Versoix, Switzerland}
\affiliation[51]{National Astronomical Observatories, Chinese Academy of Sciences, A20 Datun Road, Chaoyang District, Beijing, 100101, P.~R.~China}

\emailAdd{corentin.ravoux@clermont.in2p3.fr}

\abstract{We present the one-dimensional Lyman-$\alpha$ forest power spectrum measurement derived from the data release 1 (DR1) of the Dark Energy Spectroscopic Instrument (DESI). The measurement of the Lyman-$\alpha$ forest power spectrum along the line of sight from high-redshift quasar spectra provides information on the shape of the linear matter power spectrum, neutrino masses, and the properties of dark matter. In this work, we use a Fast Fourier Transform (FFT)-based estimator, which is validated on synthetic data in a companion paper. Compared to the FFT measurement performed on the DESI early data release, we improve the noise characterization with a cross-exposure estimator and test the robustness of our measurement using various data splits. We also refine the estimation of the uncertainties and now present an estimator for the covariance matrix of the measurement. Furthermore, we compare our results to previous high-resolution and eBOSS measurements. In another companion paper, we present the same DR1 measurement using the Quadratic Maximum Likelihood Estimator (QMLE). These two measurements are consistent with each other and constitute the most precise one-dimensional power spectrum measurement to date, while being in good agreement with results from the DESI early data release.}

\begin{document}
\maketitle
\flushbottom

\section{Introduction}
\label{sec:intro}

The Lyman-$\alpha$ (\lya) forest is a set of resonant \lya~($\lambda = 1215.67$\AA) absorption features imprinted in the spectra of background objects, such as quasars, caused by intervening neutral hydrogen overdensities in the intergalactic medium (IGM) along their lines-of-sight. From ground-based observations, the \lya~forest is measured in the redshift range $2 \lesssim z \lesssim 6$. At higher redshift, the ionization state of the IGM is such that \lya~absorption is near-complete, while the atmospheric UV cutoff constrains lower redshift observations.

Thanks to the redshift effect, the absorption features measured at different but closeby wavelengths probe separate, closeby, IGM regions: for example, at $z=2.4$, absorptions seen at $2$ \AA~separation are sourced on average by clouds of IGM separated by $\sim 1.5\,h^{-1}$~Mpc. This property makes the \lya~forest a unique probe to study the matter density field on small cosmological scales at $z > 2$~\cite{gunn_density_1965,lynds_absorption-line_1971,meiksin_physics_2009,mcquinn_evolution_2016}. 

Many summary statistics based on \lya~measurements have been used to infer properties of the IGM and the underlying matter field (e.g.~\cite{Meiksin:2000ug,delaCruz:2024cai,Lee:2014kda}). Here, we focus on $\pk$, the one-dimensional power spectrum of the fluctuations of the \lya~transmission field along lines-of-sight. The \lya~transmitted flux fraction $F(\lambda)$ is the relative IGM \lya~absorption as a function of wavelength in the observer's frame, i.e., the ratio of the measured flux to the background source's intrinsic flux, in the absence of any contaminants such as absorptions by metals or instrumental effects. The corresponding flux contrast is $\delta_F = F/\bar{F}(\lambda)-1$, and $\pk$ is the one-dimensional power spectrum of the \lya~forest signal in $\delta_F$. As a decisive advantage, $\pk$ is quite directly connected to the linear matter power spectrum at redshift $z>2$ and near-Mpc scales (e.g.~\cite{Croft:1997jf,SDSS:2004aee}). More particularly, the \lya~forest one-dimensional power spectrum is among the few cosmological observables that allow to probe such highly nonlinear scales. Hence, it offers a unique window to constrain models that impact small cosmological scales. As a consequence, $\pk$ measurements has been used to test several cosmological scenarios, such as a running of the primordial matter power spectrum, the effect of the sum of neutrino masses~\cite{Croft:1999mm,seljak_cosmological_2006,palanque-delabrouille_neutrino_2015,yeche_constraints_2017,palanque-delabrouille_hints_2020}, and several dark matter models such as warm \cite{viel_constraining_2005,viel_how_2008,viel_warm_2013,baur_lyman-alpha_2016,yeche_constraints_2017,baur_constraints_2017,palanque-delabrouille_hints_2020}, fuzzy~\cite{irsic_first_2017,armengaud_constraining_2017}, or interacting dark matter~\cite{Hooper2022}, or primordial black holes~\cite{Murgia2019}.

The Dark Energy Spectroscopic Instrument (DESI) survey~\cite{desi_collaboration_desi_2016,desi_collaboration_desi_2016-1,abareshi_overview_2022} provides an unprecedented sample of high-redshift quasar spectra from which \lya~forest information can be extracted. DESI expects to measure around one million \lya~forest 
\cite{desi_collaboration_validation_2023}, a sample five times larger than the previous SDSS survey~\cite{bourboux_completed_2020}. A major goal of DESI is to measure the three-dimensional auto-correlation of the \lya~forest, as well as its cross-correlation with quasar positions, on large scales, to derive the BAO scale and linear growth of structures at redshift $z\sim 2.3$~\cite{bourboux_completed_2020,Gordon2023,DESIBAOlya2024,DESIY12024,DESIBAOlya2025,DESIBAOY22025}. In this article, we focus on the one-dimensional correlations of the \lya~forest, which can be studied down to small scales thanks to DESI spectral resolution ($R = \Delta \lambda / \lambda \in [2,000 - 5,100]$ for wavelength ranging from $3,600$ to $9,800$ \AA). In~\cite{Ravoux2023, karacayli_optimal_2023}, we reported on the first measurements of the related power spectrum based on the DESI Early Data Release (EDR)~\cite{desi_collaboration_validation_2023,desi_collaboration_early_2023} and its first two months (M2) data sample (noted \edrmtwo~hereafter). Two different estimators were used to compute $\pk$, one based on the Fast-Fourier Transform of individual 1D spectra (FFT \cite{Ravoux2023} and noted \desiedrfft~hereafter), and one based on a Quadratic Maximum Likelihood Estimator (QMLE, \cite{karacayli_optimal_2023}). In this article, as well as companion papers~\cite{karacayli2025, karacayliravoux2025} (noted \qmleyone~and \validpaper~hereafter), we update those measurements, using the vastly larger \lya~sample included in the DESI Data Release 1 (DR1)~\cite{DESIDR12025}, built from the first year of DESI main survey. To be more explicit, the DR1 catalog contains $521,708$ \lya~forests, around five times more than previous EDR measurements, with improvement in terms of signal-to-noise ratio. In addition to the massive statistical gain, we present a series of analysis improvements and robustness tests.

Given the abundance of quasar spectra collected over the years from different instruments, several measurements of $\pk$ were performed over a range of redshifts and wavenumbers. On the one hand, with high-resolution spectra from, e.g., VLT/UVES~\cite{murphy_uves_2018}, VLT/X-shooter~\cite{lopez_xq-100_2016} or Keck~\citep{omeara_first_2015,omeara_second_2017}, $\pk$ was measured for large values of the wavenumber $k$. On the other hand, $\pk$ was estimated for smaller values of $k$, and with higher statistical precision, thanks to the massive \lya~forest samples from the first SDSS spectroscopic survey, the Baryon Oscillation Spectroscopic Survey (BOSS)\cite{palanque-delabrouille_one-dimensional_2013} and follow-up eBOSS\cite{chabanier_one-dimensional_2019} surveys. Compared to BOSS and eBOSS data, the \lya~spectra from DESI have a better spectral resolution, therefore enabling measurements of $\pk$ up to larger $k$, overlapping with the measurements derived from the high-resolution samples.

The outline of this paper is as follows. After presenting the dataset used in section~\ref{sec:data}, including the catalogs of contaminants, we describe the current implementation of the FFT method in section~\ref{sec:method}, and present a new characterization of the impact of instrumental properties, especially the noise in section~\ref{sec:instrumental}. We then discuss in section~\ref{sec:data_splits} a series of data splits and analysis variations that were used to assess the robustness of the measurement. After that, we turn to our estimate of the statistical covariance and systematic uncertainties in section~\ref{sec:uncertainties}. Finally, we present in section~\ref{sec:measurement} our measurement and compare it to high-resolution and eBOSS measurements.

\section{Data}
\label{sec:data}

\subsection{The Dark Energy Spectroscopic Instrument}

\begin{figure}
    \centering
    \includegraphics[width=\linewidth]{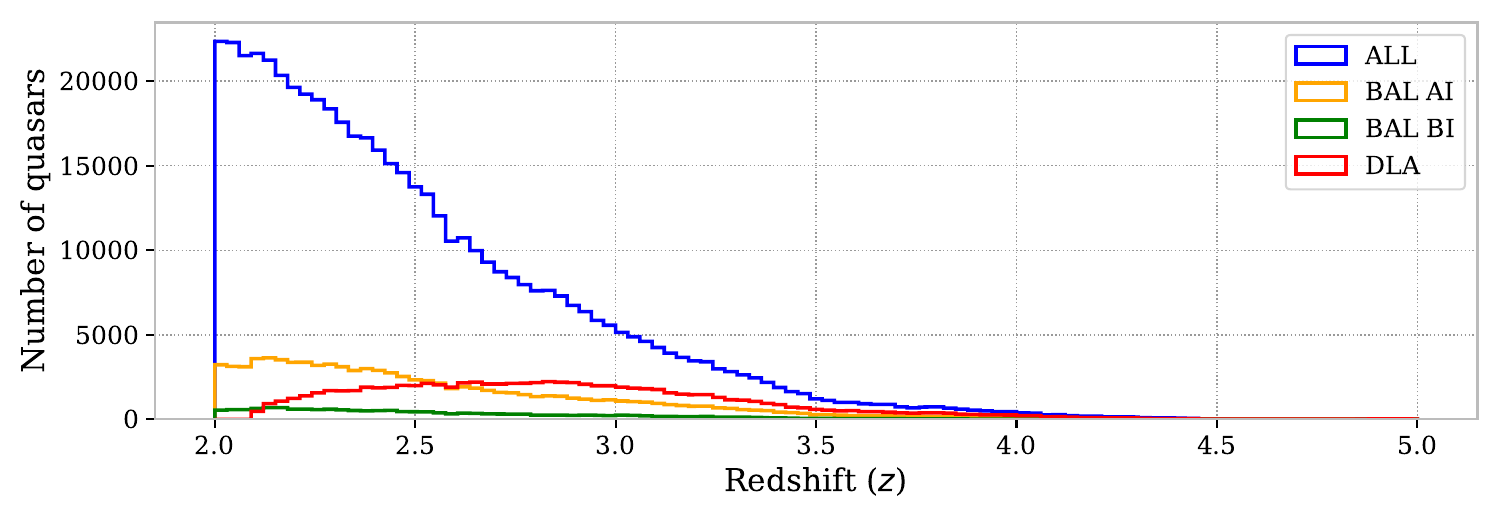}
    \caption{Histogram of the quasar redshifts in the \drone~data release. All quasars with redshift $2.0 < z < 5.0$ are represented in blue. The distribution of Broad Absorption Line (BAL) quasars defined by the Absorption Index criterion ($AI >0$) appears in yellow, and the distribution of quasars with Balnicity Index criterion ($BI>0$) is in green. The distribution of quasars containing at least one detected Damped \lya~(DLA) object in their spectra is shown in red.}
    \label{fig:quasar_redshift_histogram}
\end{figure}

The Dark Energy Spectroscopic Instrument (DESI) is a multi-object spectrograph whose purpose is to measure the spectra of more than $40$ million galactic (stars) and extra-galactic objects (galaxies and quasars). The main science goal of DESI is to infer the properties of dark energy using Baryon Acoustic Oscillations (BAO) and Redshift Space Distortions (RSD) measured in galaxy, quasars, and \lya~correlation functions. Additionally, the significant statistics and variety of observations allows DESI to cover broader scientific objectives, including the one presented in this article. The focal plane of DESI is equipped with $5,000$ robotic fiber positioners~\cite{desi_collaboration_desi_2016-1,Silber2022,Miller2023,Schlafly2023,Poppett2024} to quickly reconfigure the observation pattern and point towards pre-selected targets~\cite{dey_overview_2019}. A full description of the DESI instrument is given in~\cite{abareshi_overview_2022}. We use the first data release noted \dronefull~(or \drone~for conciseness), which results from the first year of DESI observation~\cite{DESIDR12025}.

\subsection{Quasar catalog and spectra}

Quasars candidates are pre-selected with a dedicated targeting to obtain a denser sample of high-redshift targets~\cite{dey_overview_2019,yeche_preliminary_2020,chaussidon_target_2022}. The \dronefull~quasar sample is constructed by automated classifiers. A first template fitting software called \texttt{redrock} \git{desihub/redrock}{}~\cite{Brodzeller2023} is used to classify quasars, galaxies, and stars, and to provide a robust redshift measurement. A dedicated neural network software called \texttt{QuasarNet} \git{desihub/QuasarNP}{}~\cite{busca_quasarnet_2018,farr_optimal_2020} picks up quasars missed by \texttt{redrock}. As \texttt{QuasarNet} is designed explicitly for classification, the redshift of retrieved quasars is determined by running \texttt{redrock} again with an informed prior. Finally, a \mgii~line fitter algorithm catches low-redshift quasars ($1.5 < z < 2.0$) whose redshifts were correctly measured, but which have been classified as objects of another type. The full description of the pipeline used to generate the quasar catalog is given in~\cite{chaussidon_target_2022} and in~\cite{DESIDR12025} specifically for DR1. The histogram of the \drone~quasar redshifts is represented in figure~\ref{fig:quasar_redshift_histogram}. It contains $521,708$ quasars with a redshift in the \lya~forest range of interest ($z > 2.0$). As detailed in~\cite{DESIY1sample2024}, 66.8\% of the targeted quasars were spectroscopically confirmed in DR1. Additionally, some quasars are obtained from the spectra of objects initially targeted as other classes, such as stars or galaxies (emission line galaxies in particular). A total of 8\% of quasars are added to the sample from misclassification of other classes.

Several modifications are made to obtain this quasar catalog. First, spectra from secondary target programs dedicated to the search for high-redshift quasars are added. A visual inspection campaign of those targets is carried out (see the companion paper \qmleyone~for more details). Second, we remove quasars whose measured individual power spectrum appears as an outlier. The full \drone~sample cannot be entirely visually inspected and still contains outliers passing the automated classifier criteria. We carry out a pre-study in which the individual power spectrum of each \lya~forest is measured. Taking the full \drone~catalog, we remove quasars for which the average of individual power spectrum over large scales ($k< 0.8$ \invAA) is more than $20\sigma$ higher than the average for all quasars, where $\sigma$ is the standard deviation over all quasars. This analysis is carried out in the \lya~and side-band SB1 ($1,270 < \lambda_{\mathrm{rf}} < 1,380$ \AA) regions used later in this paper. A visual inspection showed that the removed spectra are indeed quasar spectra but exhibiting spikes caused by instrumental artifacts. A small number of $7$ and $18$ quasars are removed by conducting this study in the \lya~and SB1 region, respectively. This removal eliminates spurious oscillations in the measurement of the one-dimensional power spectrum.

\subsection{Other catalogs}

The computation of one-dimensional power spectra necessitates accounting for several effects that causes changes in the spectra (absorptions or emissions), and that are not directly related to IGM absorptions. Our choice concerning these contaminants is to remove the spectra from the data sample or mask the impacted spectral region. 

First, the Broad Absorption Lines (BAL) quasars present specific blueshifted absorptions caused by the quasar torus, thus directly related to its surroundings. These specific quasars are found using the \texttt{baltools} \git{paulmartini/baltools}{} software~\cite{Filbert2023}, which is a $\chi^2$ minimizer that measures blueshifted \civ~or \siiv~absorptions compared to an unabsorbed quasar continuum. The \texttt{baltools} software computes for each quasar spectrum the Balnicity Index (BI) and Absorption Index (AI)~\cite{Filbert2023} defined as

\begin{equation}
\begin{split}
    B I &=-\int_{25000}^{3000}\left[1-\frac{\hat{f}(v)}{0.9}\right] C(v) d v\, , \\
    A I &=-\int_{25000}^0\left[1-\frac{\hat{f}(v)}{0.9}\right] C(v) d v\, ,
\end{split}
\end{equation}

\noindent where $\hat{f}$ is the normalized flux, i.e., the observed quasar flux divided by a model fitted to the quasar without BAL features, $v$ is the separation from the \civ~line center in velocity units, and $C$ is a constant null term unless  $[1 - \hat{f}(v) / 0.9]$ is greater than zero for more than $2,000\ \invkms$. The $BI > 0$ is a stricter criterion to define a BAL quasar, thus all $BI > 0$ BAL quasars are included in the $AI > 0$ sample. The $\pk$ measurement is carried out after removing all quasars that satisfies the $BI > 0$ criterion ($\sim 4 \%$ in the total data set).

Specific absorptions called High-Column Density (HCD) systems are present in the spectrum of some quasars. They are caused by the passage of the quasar light in the vicinity of a galaxy, called a circumgalactic medium. HCD objects contaminate the \lya~forest by imprinting a saturated absorption that spans a much larger spectral range than the size of the galaxy itself and adds metal absorption lines present in the circumgalactic medium \cite{mcdonald_physical_2005}. The HCD objects are detected using the combination of a convolutional neural network (CNN) algorithm \texttt{desi-dlas} \git{cosmodesi/desi-dlas}{} \citep{parks_deep_2017} and a Gaussian process (GP) algorithm \git{jibanCat/gpy_dla_detection}{} \citep{ho_damped_2021}. The GP algorithm provides a better estimate of the HCD column density $N_{\mathrm{\hi}}$, and the CNN a better detection completeness. We only consider the Damped \lya (DLA) objects, defined by $N_{\mathrm{\hi}} > 10^{20.3}~\mathrm{cm}^{-2}$. We take the same criteria as \desiedrfft~to define our sample: valid DLA detection by CNN with a confidence level higher than $0.2$ for quasars continuum-to-noise ratio $\cnr >3$ and a confidence level higher than $0.3$ when $\cnr <3$. For the GP algorithm, we take a minimal $0.9$ confidence level. The final catalog results from the merging of DLA detected by either algorithm. When both algorithms detect the DLA, the GP column density is used. The core region of a detected DLA, defined by an induced absorption larger than $20~\%$, is masked at the spectrum level. The remaining DLA damping wings are corrected at the spectrum stage using a Voigt profile (see e.g.~\cite{bourboux_completed_2020} for details regarding DLA masking). Approximately $16$ \% of the quasars in the \drone~catalog have at least one detected DLA in their spectrum. The redshift distribution of quasars presenting at least one detected DLA and that of quasars passing the $AI > 0$ or $BI>0$ BAL quasar criteria are shown in figure~\ref{fig:quasar_redshift_histogram}. 

Finally, absorption lines caused by passing through the Milky Way and atmospheric emission lines cause spurious spectra modifications even after being corrected by the DESI pipeline. We mask the lines in the line catalog adapted to DESI spectral resolution and developed in~\desiedrfft (\git{corentinravoux/p1desi/tree/main/etc/skylines/list_mask_p1d_DESI_EDR.txt}{}). 
The transmitted flux fraction of masked pixels (DLA or lines) is fixed to its average value ($F = \overline{F}$), which corresponds to a zero flux contrast. This masking induces a bias on $\pk$ that will be corrected in section~\ref{sec:measurement}.

\section{Method}
\label{sec:method}

The one-dimensional power spectrum is computed with the same method as \desiedrfft. This section aims to summarize the continuum fitting procedure and the $\pk$ FFT estimator used in this paper and to update the studies carried out in \desiedrfft~with the \drone~dataset employed here.

\subsection{Continuum fitting procedure}
\label{subsec:continuum}

The continuum fitting procedure estimates the flux contrast $\df$, i.e., the normalization of the flux variation caused by all absorptions in the \lya~forest, from the measured flux $f$ at a specific wavelength $\lambda$:

\begin{equation}
    \label{eq:delta_flux_continuum}
    \df(\lambda) =   \frac{F(\lambda)}{\overline{F}(\lambda)} - 1 =  \frac{f(\lambda)}{C_{\mathrm{q}}(\lambda,z_{\mathrm{q}})\overline{F}(\lambda)} - 1\,, 
\end{equation}

\noindent where $C_{\mathrm{q}}(\lambda,z_{\mathrm{q}})$ is the quasar continuum corresponding to the unabsorbed flux emitted by the quasar $q$ and expressed at the quasar redshift $z_q$. The term $\overline{F}$ corresponds to the average fraction of transmitted flux in the IGM. We use the \texttt{picca} \git{igmhub/picca}{} \citep{du_mas_des_bourboux_picca_2021} software package to fit for the product $C_{\mathrm{q}}(\lambda,z_{\mathrm{q}})\overline{F}(\lambda)$ using the following parameterization for the continuum:

\begin{equation}
    \label{eq:continuum_quasar}
    C_{\mathrm{q}}(\lambda,z_{\mathrm{q}}, a_{\mathrm{q}}, b_{\mathrm{q}}) = \left(a_{\mathrm{q}} + b_{\mathrm{q}}\lambda\right) C\left(\lambda_{\mathrm{rf}}\right)\,,
\end{equation}

\noindent where $\lambda_{\mathrm{rf}}=\lambda / (1+z_{\mathrm{q}})$ is the rest-frame wavelength, $a_{\mathrm{q}}$ and $b_{\mathrm{q}}$ are parameters fitted for each quasar individually, accounting for their variability, $C$ is a function used to model the common continuum of all quasars. The latter is computed as the average of all the quasar spectra available in the sample. The fitting procedure is done iteratively for the ensemble of all quasars, updating the common continuum and fitting for the $a_{\mathrm{q}}$ and $b_{\mathrm{q}}$ terms at each iteration, by minimizing the following log-likelihood:

\begin{equation}
\label{eq:likelihood}
\ln \mathcal{L} = - \frac{1}{2} \sum_{i} \frac{\left[f (\lambda_{i})-\overline{F}(\lambda_{i}) 
C_{\mathrm{q}}\left(\lambda_{i},z_{\mathrm{q}}, a_{\mathrm{q}}, b_{\mathrm{q}}\right)\right]^{2}}{\left(\overline{F}(\lambda_i) C_{\mathrm{q}}\left(\lambda_{i},z_{\mathrm{q}}, a_{\mathrm{q}}, b_{\mathrm{q}}\right)\right)^{2}}-\ln \left[\left(\overline{F}(\lambda_i) C_{\mathrm{q}}\left(\lambda_{i},z_{\mathrm{q}}, a_{\mathrm{q}}, b_{\mathrm{q}}\right)\right)^{2}\right]\,.
\end{equation}

The noise associated to each flux contrast is directly given by the estimate of the DESI pipeline noise, noted $\sigma_{\mathrm{pip,q}}(\lambda)$ (see~\cite{guy_spectroscopic_2022} and \desiedrfft), normalized by the $C_{\mathrm{q}}(\lambda,z_{\mathrm{q}})\overline{F}(\lambda)$ product. We then define the mean signal-to-noise ratio of a quasar $q$ as

\begin{equation}
\label{eq:mean_snr}
\overline{\mathrm{SNR}} = \left\langle \frac{f(\lambda)}{\sigma_{\mathrm{pip,q}}(\lambda)} \right\rangle_{\lambda}\,.
\end{equation}

Following \desiedrfft, we only keep quasars with $\snr > 1$. The fitting procedure is performed for an observed wavelength range of $3,600 < \lambda < 7,600$ \AA and a rest-frame range of $1,050 < \lambda_{\mathrm{rf}} < 1,180$ \AA. The quasar spectra are linearly binned in observed wavelength with $\Delta \lambda_{\mathrm{pix}} = 0.8$ \AA. We adopt the same rebinning scheme of the common continuum $C$ in the rest-frame basis with a bin size equal to $\Delta \lambda_{\mathrm{pix,rf}} = 2.67$ \AA. Finally, the stack of all flux contrasts is set to zero at the end of the procedure.

\subsection{Fast Fourier Transform estimator}
\label{subsec:fft}

The one-dimensional \lya~power spectrum, noted $\pk$, is defined as the power spectrum of the \lya~absorption contrast $\dlya$ along the quasar line-of-sight. Following \desiedrfft, we include in the $\pk$ definition the cross-correlations between the \lya~absorption line and other IGM elements whose rest-frame absorption is very close to the \lya~line. In particular, we include absorptions from \siii~and \siiii~($\lambda_{\mathrm{\siii}} = 1,190$ and $1,193$ \AA, and $\lambda_{\mathrm{\siiii}} = 1,206.50$ \AA) whose cross-correlation with \lya~absorption is modeled at the cosmological interpretation stage. Neglecting the \siii~and \siiii~auto-correlations as they correspond to minor metal lines, we mathematically define $\pk$ as

\begin{equation}
    \begin{split}
    \label{eq:p1d_def}
    (2\pi) \dD(k-k') \pk(k) = & \left\langle\dlya(k) \dlya^{*}(k') \right \rangle + \left\langle\dlya(k)\dsiii^{*}(k)\right \rangle + \left\langle\dlya^{*}(k)\dsiii(k)\right \rangle \\
    & + \left\langle\dlya(k)\dsiiii^{*}(k)\right \rangle + \left\langle\dlya^{*}(k)\dsiiii(k)\right \rangle \,,
\end{split}
\end{equation}

\noindent where $\dD$ is the Dirac function, $\dsiii$ and $\dsiiii$ are the absorption contrasts associated to the \siii~and \siiii~lines.

To compute $\pk$, we first separate the flux contrast $\df$ from each spectrum into three non-overlapping sub-forests of equal size, $L_{\mathrm{sub}} = 43.3$ \AA\ in the rest frame, with a flux contrast noted $\dfs$. Each of the $3 N_\mathrm{q}$ sub-forests (with $N_\mathrm{q}$ the total number of quasars) is referred to with the index $s$ in the following equations. This splitting reduces correlations between redshift bins and increases the smallest accessible wavenumber to $k_{\mathrm{min}} = 2\pi / (L_{\mathrm{sub}} (1 + z_{\mathrm{min}}))  = 0.0468$ \invAA. Furthermore, the resulting sub-forest covers at most $\Delta z = 0.2$, and we use this value to define the redshift binning for $\pk$. The redshift associated with a sub-forest is determined by the center of its observed wavelength range, which assigns it to a redshift bin. Therefore, each redshift bin contains information from adjacent redshift bins, as sub-forests with a central redshift near a bin edge have half their spectrum extending beyond their redshift bin. It implies the existence of an inter-redshift correlation, which is not accounted for in this work but will be in future studies. The sub-forest flux contrasts are then transformed to Fourier space by \texttt{picca} using a multiprocessed Fast Fourier Transform (FFT) algorithm.

The $\pk$ FFT estimator is built under the main assumption that each sub-forest in our dataset provides an independent realization of the IGM \lya~absorptions at the redshift at which they occur. In other words, we assume that the average of all sub-forests gives an unbiased estimator of the one-dimensional power spectrum, and we neglect correlations between sub-forests. The FFT estimator is derived from the relation between the observed flux contrast $\dfs$ obtained after continuum fitting and the \lya~contrast $\dlyas$. This relation is derived by considering all the instrumental (noise, resolution) and astrophysical (metals) effects that imprint absorptions in the \lya~region. Following the decomposition detailed in \desiedrfft, the FFT $\pk$ estimator for the wavenumber bin $A$ and redshift bin $z$ is defined by

\begin{equation}
\label{eq:p1d_estimator}
\pk(A,z)  = \left\langle\frac{\prs(k)-\pns(k)}{\rmats^{2}(k)} \right\rangle_{s \in z, k\in A} -\pme(A,z) = \left\langle \ps(k) \right\rangle_{s \in z, k\in A} - \pme(A,z)\,,
\end{equation}

\noindent where $\prs$ is the power spectrum of the flux contrast $\dfs$, $\pns$ is the noise power spectrum, $\pme$ is the metal power spectrum resulting from the combination of metal forests whose emission lines are located redwards of the three \lya, \siii~and \siiii~absorptions, $\ps$ is a practical term called individual power spectrum of the sub-forest $s$, and $\rmats$ is the average of the Fourier transform of the resolution matrix provided by the DESI pipeline. This matrix encodes for each spectrum and each wavelength the complete resolution information, resulting from spectroscopic extraction (see e.g.~\citep{bolton_spectro-perfectionism_2010,guy_spectroscopic_2022} and \desiedrfft\ for more details). The resolution matrix is symmetric and is averaged for each spectra pixel in the wavelength space before applying the Fourier transform. We call $\prs$, $\pns$ and $\ps$ power spectra although they correspond to one single sub-forest and only their average over many sub-forest is a measurement of the power spectrum.

The \texttt{picca} software is used to compute this FFT estimator. In agreement with \desiedrfft, we fix our minimal redshift to $z_{\mathrm{min}} = 2.1$ to avoid the $\lambda \lesssim 3,700$ \AA\ region of the spectra where atmospheric absorptions significantly increase the flux noise. As we apply several masks (DLA, BAL, and atmospheric emission line catalogs), we remove sub-forests shorter than 75 wavelength pixels or with more than 120 masked pixels. We compute the raw power spectrum for each sub-forest directly from the flux contrast as $\prs = \left| \dfs(k)^2\right|$. The noise power spectrum $\pns$ is computed using the DESI pipeline noise. The software generates a large number ($N_{\mathrm{G}} = 2500$) of Gaussian signals $\dnoises$ with standard deviation equal to $\sigma_{\mathrm{pip,s}}(\lambda)$ for each wavelength, and the noise power spectrum is obtained by averaging those signals in Fourier space, i.e.\ $\pns(k) = \left\langle \left|\dnoises\right|^2 \right \rangle_{N_{\mathrm{G}}}$.

The \texttt{picca} software averages the individual power spectra with a $\snr$ weighting scheme. In each bin $(A,z)$, we distribute the power spectra of individual sub-forests in bins in the signal-to-noise ratio of the sub-forests $\snr_s$. We compute the variance of the power spectra in each $\snr_s$ bin. We then fit these variances with a simple least-squares minimization as a function of $\snr_s$ according to:

\begin{equation}
    \sigma_{(A,z)}^2\left(\snr_s\right)=\frac{a_{A,z}}{\left(\snr_s-1\right)^2}+b_{A,z}\, .
\end{equation}

The fitted terms $a_{A,z}$ and $b_{A,z}$ are used to define the inverse-variance weights of each sub-forest as $w_{A,s} = 1/ \sigma_{(A,z)}^2\left(\snr_{s}\right)$. By construction, the weights tend to zero when $\snr$ tends to 1, in agreement with the applied $\snr > 1$ cut. Finally, the weighted average in equation \ref{eq:p1d_estimator} is explicitly computed as: 

\begin{equation}
\label{eq:p1d_averaging}
\left\langle \ps(k) \right\rangle_{s \in z, k\in A} = \frac{\sum_{s \in z, k\in A} w_{A,s} \ps (k) }{\sum_{s \in z, k\in A} w_{A,s} }\, .
\end{equation}

\section{Instrumental characterization of one-dimensional power spectrum}
\label{sec:instrumental}

\subsection{Updates on \drone~data set}
\label{subsec:drone_update}

Some methods applied to the analysis of \edrmtwo~DESI datasets in \desiedrfft~are applied in this paper to the \drone~data set without any modifications. We review the new characterization of those instrumental features with the \drone~data set. The updated figures are located in appendix~\ref{appendix:updated_plots}.

The metal power spectrum, $\pme$ in equation~\ref{eq:p1d_estimator}, is estimated using the side-band region SB1 ($1,270 < \lambda_{\mathrm{rf}} < 1,380$ \AA). We show the SB1 power spectrum in figure \ref{fig:metal_power}, which is measured using the averaging of equation~\ref{eq:p1d_estimator} in the SB1 rest-frame range. Here, we use a rest-frame range distinct from the \lya~forest one, but with the same observed wavelength range. The quasars selected for SB1 are then located at lower redshifts than those selected for \lya. Consequently, our correction assumes that the continuum fitting process can catch the quasar emission variability as a function of redshift. Testing this hypothesis would require the full modeling of metal forests in the side-band region of the mocks, which is out of the scope of this paper.

As in~\desiedrfft, we average the side-band power spectrum over redshift bin on a common rest-frame wavenumber binning $k_{\mathrm{rest}}=(1+z) \times k_{\mathrm{obs}}$. This average is fitted with a least-square minimization by a physically motivated model considering \siiv~and \civ~doublets oscillations:

\begin{equation}
    \label{eq:side_band_model}
    \psbm(k) = C \times k^{-\epsilon} + \sum_{i \in [\siiv, \civ]} C_{i} e^{-c_{i} k}  \sin\left(2\pi \left(\frac{k}{k_{i}}\right) + \psi_{i} \right)\,.
\end{equation}

The evolution of the SB1 power spectrum as a function of redshift is accounted for in a second step, by multiplying and fitting a linear function independently for each redshift, $\psbmz(k) = (a_z k +b_z) \psbm(k)$. The resulting fitted functions are used as the metal power spectrum correction in equation \ref{eq:p1d_estimator}. The fitted parameters for SB1 power spectrum are similar to the \edrmtwo\ ones in \desiedrfft. The \drone~power spectrum contains a larger number of sub-forests (by at least a factor $5$ for all redshifts) and consequently gives lower error bars.

The \desiedrfft~method is applied to measure the resolution damping with \drone~dataset. The resolution damping term in equation~\ref{eq:p1d_estimator} is the average of the Fourier transform of the resolution matrix. The average resolution damping is shown in Fig.~\ref{fig:resolution_characterization} and is almost identical to the one measured for \edrmtwo~data set in \desiedrfft. Consequently, we choose the same maximal wavenumber value $k_{\mathrm{max}} = 2.0$ \invAA, corresponding to a factor 5 correction.

\subsection{Improving noise assessment}
\label{subsec:noise}

\begin{figure}
    \centering
    \includegraphics[width=\linewidth]{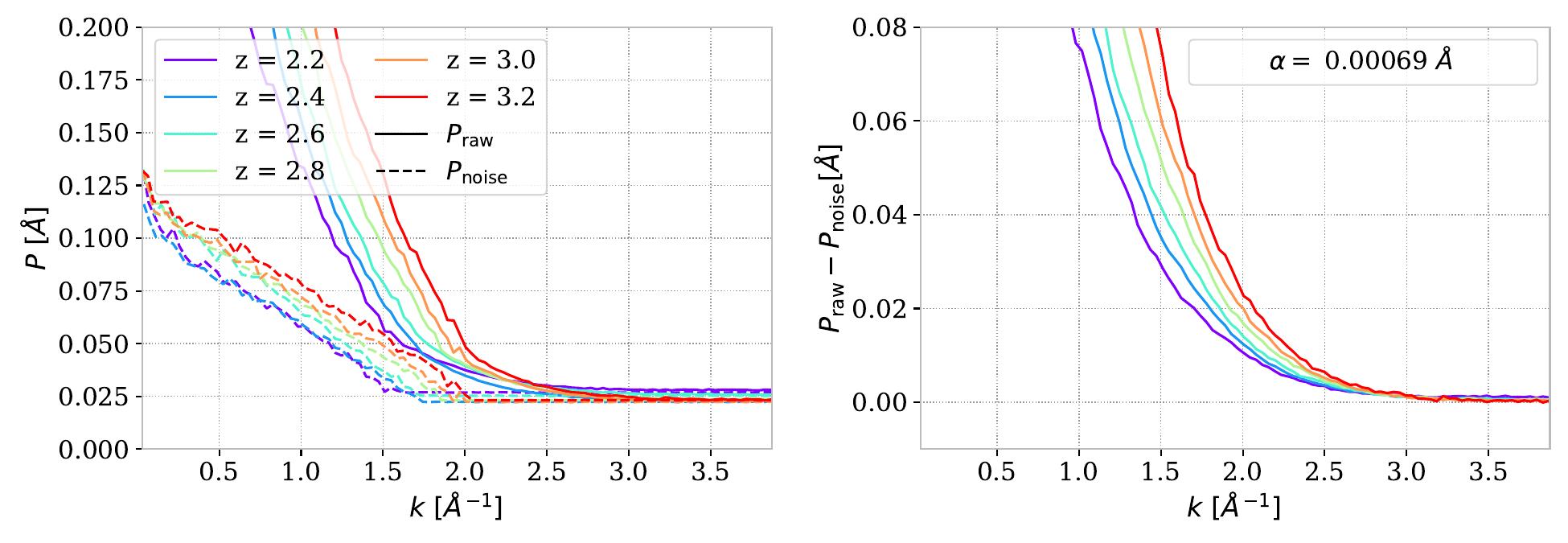}
    \caption{(left) Average raw ($\pr$) and noise ($\pn$) power spectra for the \drone~dataset as a function of the wavenumber expressed in \invAA. (right) Difference between raw and noise power spectra on the same dataset and resulting asymptotic difference $\alpha$.}
    \label{fig:noise_power}
\end{figure}

We measure the noise power spectrum with the method detailed in section \ref{subsec:fft} directly from the DESI pipeline output. As highlighted in~\desiedrfft, characterizing the noise sources in DESI is challenging, and the noise level can be underestimated. To have an alternative way to estimate the noise level, we use the smallest scales of $\pk$, for which the resolution damping completely suppresses the signal, resulting in an expected equality between the raw and noise power spectra. Consequently, the measurement of $\prs$ at the largest wavenumber is a suitable noise level estimator. The noise and raw power spectra are compared in figure~\ref{fig:noise_power} for the \drone~dataset. 

We derive an additive noise correction by averaging the $\prs - \pns$ difference for wavenumber where the resolution damping is higher than 98 \% ($k_{\mathrm{res},98} = 3.1$ \invAA), which gives $\alpha = \left \langle \prs(k) \right \rangle_{\forall z, k > k_{\mathrm{res},98}} - \left \langle \pns(k) \right \rangle_{\forall z, k > k_{\mathrm{res},98}}$. The measured $\alpha = 6.9 \times 10^{-4}$ \AA~for \drone~corresponds to a 2.8 \% noise power spectrum underestimation. This is lower than the value $\alpha = 1.09 \times 10^{-3}$ obtained for DESI-M2 in \desiedrfft, a dataset with similar properties, which means we have better control over the noise characterization thanks to the dataset quality. In contrast to~\desiedrfft, the noise power spectrum is not flat for $k \lesssim 2.0$ \invAA. This is due to the $\snr$ weighting used for \drone\ and not for \desiedrfft. Indeed, the noise power spectrum is flat for all individual power spectra, but since we are considering different weights in the $k \in[0.0 - 2.0]$ \invAA~range, the $\snr$ weights distribution impacts the averaging of the noise power spectrum in a wavenumber-dependent way.

To control the noise power spectrum even better, we developed an alternative power spectrum estimator called cross-exposure, which does not need to model or measure the noise power spectrum explicitly. The DESI standard pipeline coadds the multiple exposures of a given quasar to create less noisy spectra~\cite{guy_spectroscopic_2022}. In contrast, the cross-exposure estimator computes the one-dimensional individual cross-power spectrum between all distinct exposures of a given quasar. This provides an estimator free from noise influence if the noise realizations are independent between exposures. 

Therefore, to minimize the impact of correlated noise sources, we select exposures to get only one per quasar and per night. It removes the noise sources resulting from exposures using the same calibration images, such as bias and flat calibrations, but not necessarily that resulting from dark calibration, which is not performed each night (see e.g.~\cite{guy_spectroscopic_2022} or appendix C of \desiedrfft~for an extensive description of noise sources). 

For a quasar with $N_{\mathrm{exp}}$ exposures in DESI, we define the following cross-exposure operator:

\begin{equation}
    \mathcal{P}_X\left[\vecdone, \vecdtwo\right]=\frac{1}{N_{\mathrm{exp}}\left(N_{\mathrm{exp}}-1\right)} \sum_m^{N_{\mathrm{exp}}} \sum_{n \neq m}^{N_{\mathrm{exp}}}\left|\done^{m}(k) \dtwo^{n}(k)^*\right| \, ,
\end{equation}

\noindent where $\vecdone$ and $\vecdtwo$ are two contrast sets of size $N_{\mathrm{exp}}$ representing the contrast for different exposures. For example, $\vecdone$ can be a vector of $N_{\mathrm{exp}}$ \lya~contrasts and $\vecdtwo$ a vector of $N_{\mathrm{exp}}$ noise contrasts. For this section only, the vector notation $\vec{\cdot}$ refers to a set of contrasts for different exposures.

We replace the simple raw power spectrum coadd estimator given in section \ref{subsec:fft} by the cross-exposure estimator $\prs(k) = \mathcal{P}_X \left[\vecdfs(k),\vecdfs(k)\right]$, where $\vecdfs$ is the set of sub-forest flux contrast for the $N_{\mathrm{exp}}$ exposures.

Noise correction and resolution damping must be accounted for to build a one-dimensional power spectrum estimator. Following \desiedrfft, the flux contrast can be decomposed in Fourier space as $\vecdfs(k) = \left(\vecdlyas(k) + \vecdnoises(k)\right) \times \vecrmats(k) + \vecdme(k)$. Following the demonstration in \desiedrfft~to build the $\pk$ estimator, the cross-exposure estimator $\pkcross$ yields: 

\begin{equation}
\begin{split}
    \pkcross(A,z) &= \left\langle \mathcal{P}_X\left[ \overrightarrow{\frac{\dfs(k)}{\rmats(k)}}, \overrightarrow{\frac{\dfs(k)}{\rmats(k)}} \right] + 2 \mathcal{P}_X\left[ \overrightarrow{\frac{\dfs(k)}{\rmats(k)}}, \overrightarrow{\frac{\dnoises(k)}{\rmats(k)}} \right] \right. \\
    &+ \left. \mathcal{P}_X\left[ \overrightarrow{\frac{\dnoises(k)}{\rmats(k)}}, \overrightarrow{\frac{\dnoises(k)}{\rmats(k)}} \right]   \right\rangle_{s \in z, k\in A} - \pme(A,z)\, .    
\end{split}
\end{equation}

In this decomposition, since the noise contrasts are purely random Gaussian signals, they are all independent from each other and from \lya~contrasts. Consequently, the last two terms of the decomposition are null. We numerically verified this assumption on the \drone~data set by assessing that the average of the last two terms is consistent with zero. Consequently, the cross-exposure estimator is given by:

\begin{equation}
    \label{eq:crossexp}
    \pkcross(A,z) = \left\langle \mathcal{P}_X\left[ \overrightarrow{\frac{\dfs(k)}{\rmats(k)}}, \overrightarrow{\frac{\dfs(k)}{\rmats(k)}} \right]\right\rangle_{s \in z, k\in A} - \pme(A,z)\, .    
\end{equation}

We use a new implementation in \texttt{picca} to compute this cross-exposure estimator by performing the continuum fitting on each exposure of all the \drone~quasar catalog. Since distinct exposures of the same quasar have different $\snr$, we perform the continuum fitting independently, as if each exposure were an independent quasar. We tried alternative continuum fitting methods and concluded that using the continuum fitted on the coadd of all exposures or on the exposure with the highest $\snr$ introduces biases and spurious oscillations in $\pkcross$.

\begin{figure}
    \centering
    \includegraphics[width=\linewidth]{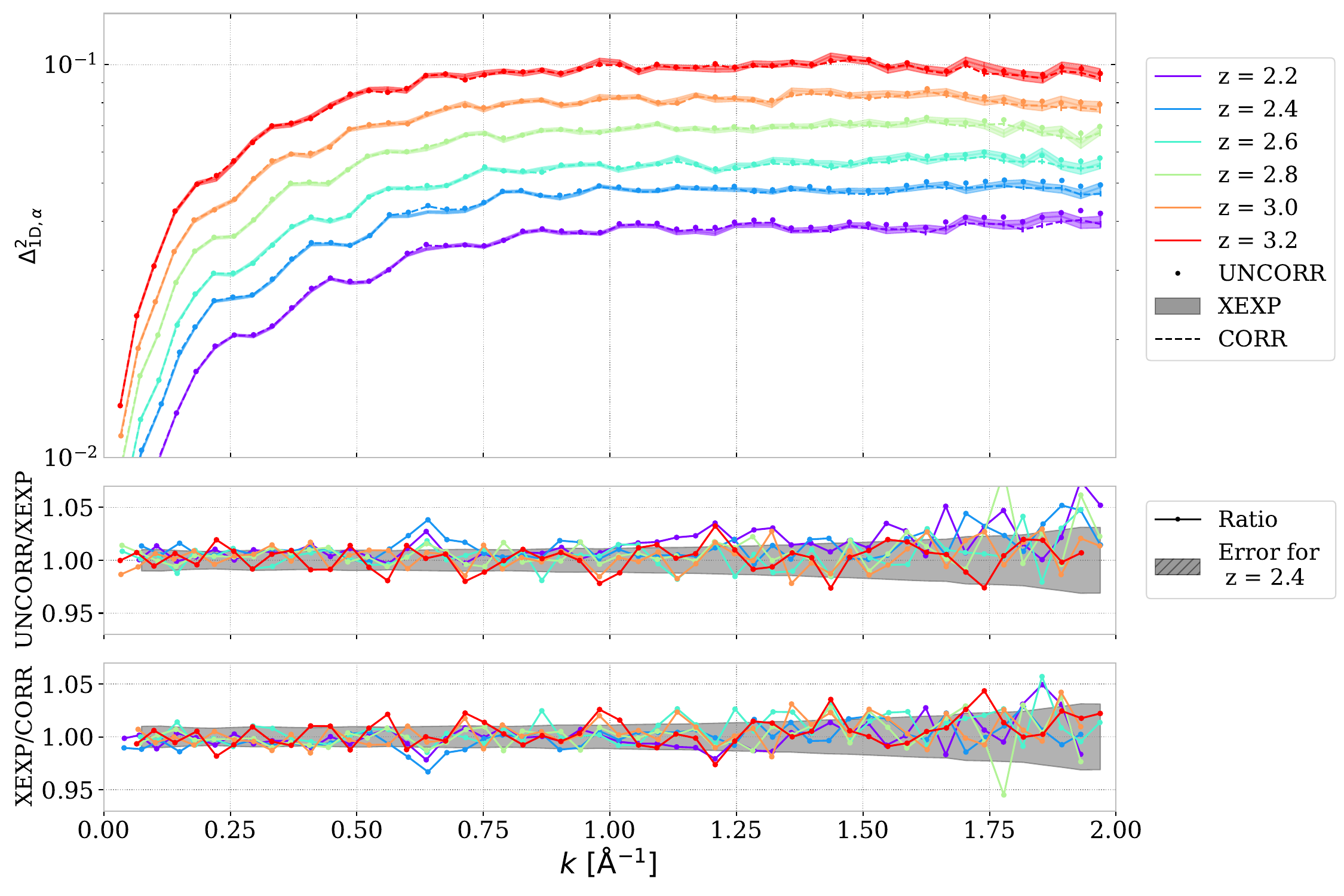}
    \caption{Comparison between the power spectrum without noise correction (UNCORR), the noise-corrected power spectrum corrected by subtracting the $\alpha$ term (CORR), and the power spectrum obtained with the cross-exposure estimator defined by equation~\ref{eq:crossexp} (XEXP). The normalized one-dimensional power spectrum ($\Delta_{1\mathrm{D},\alpha}(k) = k \pk/\pi$) is shown on the top panel, and the ratio between them is shown on the two others. For this comparison, all $\pk$ have no corrections or metal power spectrum subtraction, and only DLA and atmospheric emission lines are masked. Only the first six redshift bins are shown for clarity. For the same reason, the error bar associated with only one redshift bin ($z=2.4$) is shown in the ratios comparing the three measurements.}
    \label{fig:cross_exposure}
\end{figure}

Figure~\ref{fig:cross_exposure} compares the one-dimensional power spectrum measured on \drone~dataset before noise correction with $\pk$ after noise correction, and with the cross-exposure estimation, $\pkcross$. The two bottom panels are showing the ratios between those three analysis variations. At large wavenumber ($k \gtrsim 1.5$ \invAA), the cross-exposure estimator has the same trend as the noise-corrected measurement, indicating that both noise estimations match. The uncorrected $\pk$ have a discrepancy at small redshifts for those scales, even if the difference is still within the statistical error bar. Near $k \sim 0.6$ \invAA, both uncorrected and corrected measurements show a slight discrepancy with the cross-exposure estimator. This bump is caused by a pattern in a specific faulty DESI amplifier and is studied in detail in section~\ref{subsec:amp_split}. The cross-exposure estimator can remove part of this contamination at low redshifts.

The cross-exposure estimator only uses between $35$ and $39$ \% (depending on the redshift bin) of the quasar catalog because of the need for several exposures and the fact that we select exposures from different nights. Consequently, we do not directly use the cross-exposure estimator for high redshifts, where the statistical errors are large. 
As discussed in \ref{sec:measurement}, to better control the noise estimate, we use the cross-exposure estimator as the baseline for the redshift bins where we see small-scale and amplifier discrepancies, i.e. $z= 2.2,$ $2.4$, and $2.6$. 
For the other bins, we use the noise-corrected measurement as it agrees with the cross-exposure but results in smaller error bars due to the larger number of sub-forests and exposures available. 
Additionally, considering the extensive studies we performed here and in \desiedrfft, we assert here that we have sufficient control over the noise level to drop the associated systematic error bar, which was already overly conservative in \desiedrfft. We assume that the increase in the statistical error at low redshift due to using the cross-exposure estimator makes the systematic error due to noise negligible.

\section{Data splits and variations}
\label{sec:data_splits}

We aim at improving the robustness of the $\pk$ measurement by testing most of the hypotheses and choices we performed regarding the treatment of quasar spectra, the catalogs used (see section~\ref{sec:data}), and the methodology itself developed in section~\ref{sec:method}. As it is now standard with other \lya~analyses~\cite{DESIBAOlya2024,DESIBAOlya2025}, we vary several analysis options and perform data splitting concerning different types of properties, and compare the resulting $\pk$ to a baseline. This first application of data variations and splits for $\pk$~also aims to set the basis for this measurement. It will be improved in future studies, notably by performing it for the final baseline analysis and covariance matrix and considering additional effects. To improve readability, we explicitly label on the plot the tests that we intuitively expect to fail or pass.

For the sake of simplicity, we take a simple baseline, which uses the same methodology as the \edrmtwo\ measurement in \desiedrfft, without any correction. To be more explicit, DLAs and atmospheric emission lines are masked, and the BAL quasars verifying the $BI > 0$ criterion are removed. We do not apply noise correction, cross-exposure estimator, metal subtraction, or multiplicative correction from synthetic data. Except for cross-exposure, all the corrections that will be applied to our measurement in section~\ref{sec:measurement} are the same for all analysis variations and data splits performed here. Considering that cross-exposure only slightly impacts $\pk$, we can safely suppose that all the tests performed in this section are also valid for the final measurement for which all corrections are applied. 

This section aims to conduct a blind exploration of four data variations (DLA and BAL catalogs, continuum fitting, and averaging parameters) and four data splits (number of exposures, amplifiers, sky regions, and \civ~equivalent width).

The ratio between the baseline and the variation or split power spectra is noted $r(A,z) = P_{\mathrm{BASE}}(A,z)/P_{\mathrm{TEST}}(A,z)$. The statistical uncertainty on this ratio depends on the correlation between the two power spectra. In the data split case, we have a sub-sample of the baseline, we can compute the correlation and obtain~\footnote{This is coming from $\left( \frac{\delta (a/b)}{a/b} \right)^2 = \left( \frac{\delta a}{a} \right)^2 + \left( \frac{\delta b}{b} \right)^2 - \frac{\cov(ab)}{ab}$ and $\cov(P_{\mathrm{TEST}},P_{\mathrm{BASE}})=\var(P_{\mathrm{BASE}})$.}:

\begin{equation}
    \label{eq:err_ratio}
    \sigma_r(A,z)^2 = r(A,z)^2 \left[ \left(\frac{\sigma_{P_{\mathrm{BASE}}}(A,z)}{P_{\mathrm{BASE}}(A,z)}\right)^2 + \left(\frac{\sigma_{P_{\mathrm{TEST}}}(A,z)}{P_{\mathrm{TEST}}(A,z)}\right)^2 - 2\frac{\left(\sigma_{P_{\mathrm{BASE}}}(A,z)\right)^2}{P_{\mathrm{BASE}}(A,z)P_{\mathrm{TEST}}(A,z)} \right]\ \,
\end{equation}

\noindent where $\sigma_{P_{\mathrm{BASE}}}$ and $\sigma_{P_{\mathrm{TEST}}}$ are the standard deviations of the two measurements, see section~\ref{subsec:stat}. We then compute a $p$-value given from the $\chi^2$ cumulative distribution function $F_{\chi^2}$, and note it hereafter ``$\chi^2$ $p$-value":
\begin{equation}
    p = 1 - F_{\chi^2}\left( \sum_{(A,z)} \left[\frac{(r(A,z)-1)}{\sigma_r(A,z)}\right]^2 \right)\, .
\end{equation}
We cannot easily compute the error bars on the ratio for analysis variations; thus, they are not shown in the following. We design other statistical tests to quantify the ratio positioning and the trend as a function of wavenumber. The first test characterizes the distribution of the ratio with respect to unity. We consider the fraction of pixels below and above unity and associate a simple ``Distribution $p$-value" equal to twice the lower of the two fractions.

To characterize a potential trend with respect to the wavenumber, we want to compute the Pearson sample correlation coefficient $R_p$ between the wavenumber $k$ and the ratio $r$. However, the error on $r$ varies with $k$. In order to have a quantity whose error does not depend on $k$, we compute for each redshift the correlation between $k$ and $\left(r - \langle r\rangle\right) / \sigma_r$, where $\sigma_r$ is given by equation \ref{eq:err_ratio} without the last covariance term. It is largely overestimating $\sigma_r$, but only the variation with $k$ matters here, and we assume that we should get a fair estimate of this variation. The associated ``Correlation $p$-value" to test the existence of a correlation is given by the two-sided Student $t$-distribution:

\begin{equation}
    p = 2 P_{\mathrm{S}}\left(X \geq \left|R_p \sqrt{\frac{N-2}{1-R_p^2}}\right|\right)
\end{equation}

\noindent where $N$ is the number of data points. This $p$-value is computed for all redshifts independently, and we report the minimal one.

Some of the tests performed in this section are used to see the impact on $\pk$, and some to compare to a purer sample, which removes specific contaminants but decreases the statistics. We use the $p>0.05$ criterion to check if the test is passed in the latter case. This study allows us to conduct a data-driven exploration of potential analysis issues and modify our baseline accordingly.

\subsection{DLA and BAL catalog variations}
\label{subsec:dla_bal_split}

\begin{figure}
    \centering
    \includegraphics[width=0.49\columnwidth]{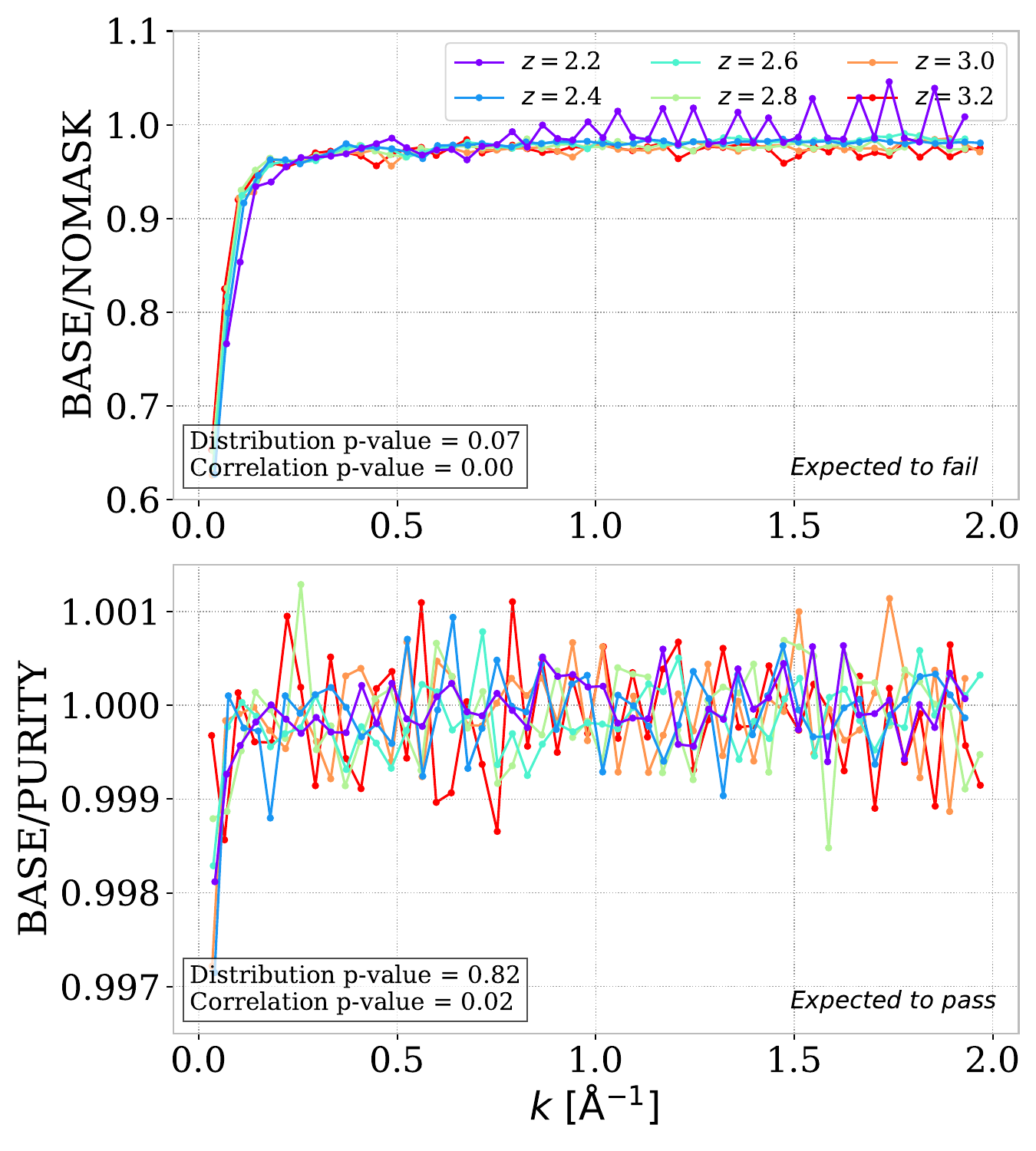}
    \includegraphics[width=0.49\columnwidth]{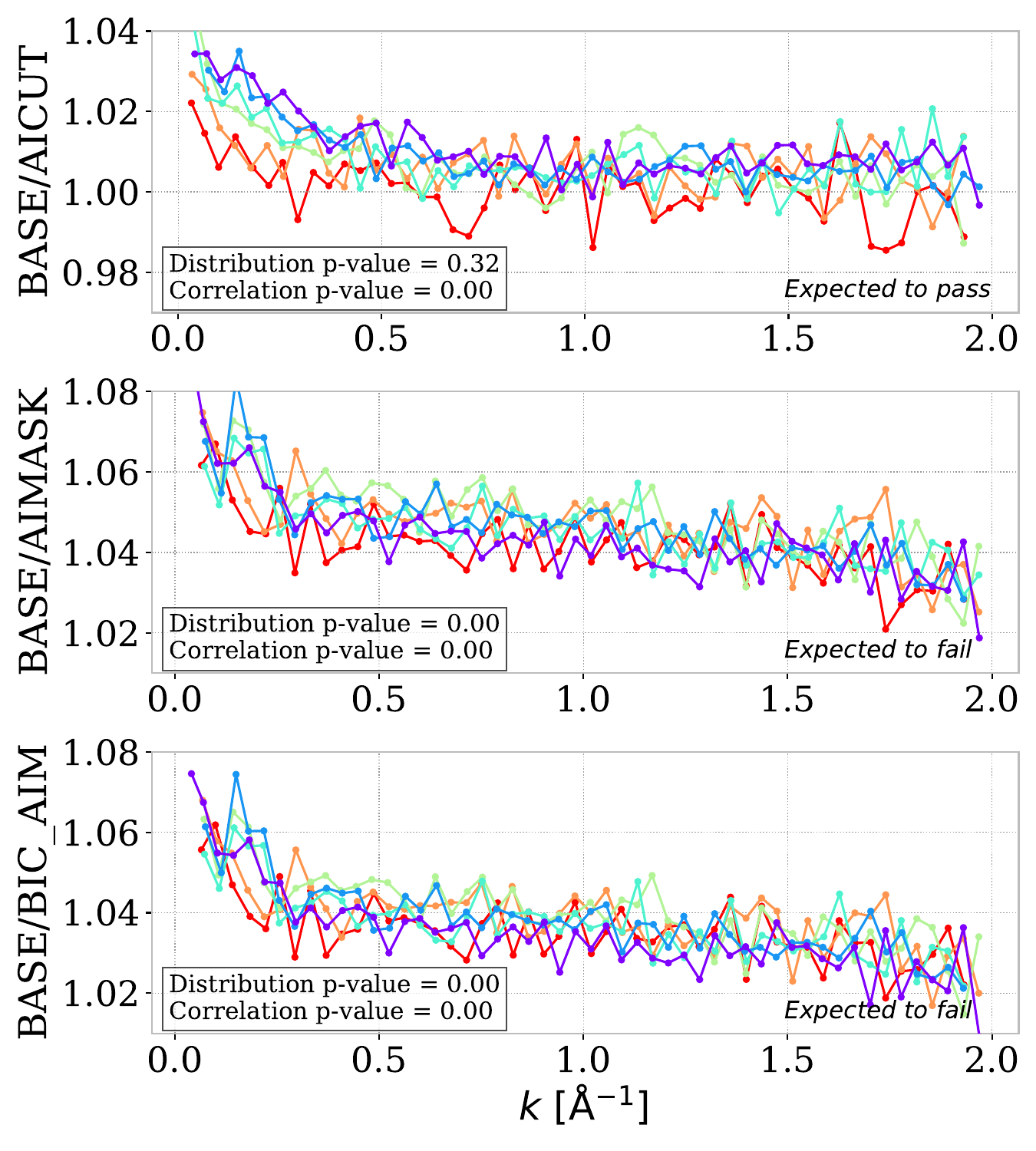}
    \caption{Ratios of the baseline power spectrum (BASE) to the power spectrum obtained from analysis variations. The baseline $\pk$ is the uncorrected power spectrum without metal subtraction. (left) Variation of the DLA treatment: without DLA masked in calculating $\pk$ (NOMASK) and with a high-purity catalog (PURITY). The level of failure in the PURITY test is well below the DLA completeness systematics associated with the final measurement in section~\ref{subsec:syst} (right) Variation of BAL treatment. The baseline removes the BAL quasar spectra detected with the $BI > 0$ criterion. The variations considered are removing the BAL quasar spectra with the $AI > 0$ criterion (AICUT), not removing any quasars from the data set, and masking all BAL absorption with the $AI > 0$ criterion (AIMASK), and removing $BI$ BAL quasar and masking $AI$ BAL absorptions (BIC\_AIM). The failure of the AICUT test shows that AI BAL significantly affect the baseline, motivating us to change the baseline.}
    \label{fig:dla_bal_splits}
\end{figure}

The DLA catalog used for masking $\pk$ was created with the specific CNN and GP confidence cuts in section~\ref{sec:method}. Variations of those parameters provide insight into DLA completeness. The left panel of figure~\ref{fig:dla_bal_splits} shows the two variations related to the DLA catalog. We start by not masking DLA (NOMASK) and comparing it to the baseline. In agreement with the previous measurement in \desiedrfft~or on simulation in~\cite{rogers_simulating_2017}, not masking DLAs greatly increases the power spectrum at the largest scales considered. This test is used in section \ref{subsec:syst} to derive a systematic uncertainty associated with the DLA completeness.

Conjointly with the companion paper \qmleyone, we modify the confidence cut to obtain a higher purity version of the DLA catalog. This alternative cutting uses the same confidence cut ($CL >0.9$) for GP, and a continuum-to-noise ratio ($\cnr$)-dependent cut for the CNN: $CL > 0.2$ for $\cnr > 3$, $CL > 0.4$ for $2 < \cnr < 3$, and $CL > 0.5$ for $\cnr < 1$. The resulting catalog contains approximately $31$ \% less DLA than the high-completeness version used for the baseline. When using this high-purity catalog, the obtained power spectrum is compared to the baseline in the last panel of Figure~\ref{fig:dla_bal_splits} (PURITY). The $p$-values reported show a statistically significant correlation with respect to wavenumber, coming from small wavenumbers and small redshifts. We note, however, that this difference is small (around 2$\%$) and is well below the level of the systematic uncertainties that we derive in section~\ref{subsec:syst} for DLA completeness. Thus, employing a purer or more complete sample of DLA does not largely impact our measurement. Furthermore, even if the two catalogs give different DLA numbers, the impact of masking is not significantly changed. We choose to keep the high-completeness catalog as the baseline.

\begin{figure}
    \centering
    \includegraphics[width=0.7\columnwidth]{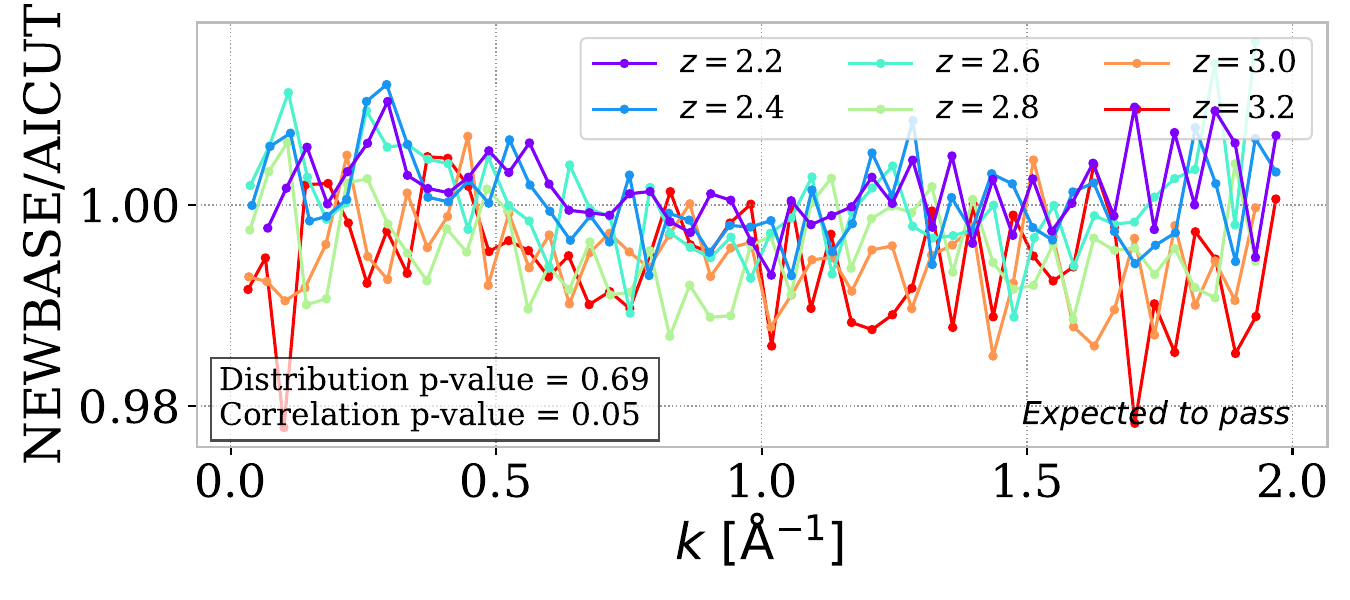}
    \caption{Ratios of the new baseline power spectrum (NEWBASE) with the AICUT BAL variation of figure \ref{fig:dla_bal_splits} for which all BAL quasars passing the $AI > 0$ criterion are removed. The new baseline removes $BI$ BAL quasars, masks $AI$ BAL absorption, and is corrected from the effect of masking with correction derived in \validpaper.}
    \label{fig:new_bal_split}
\end{figure}

We have a larger number of sub-forests than in \desiedrfft, sufficient to study the effect of the completeness of BAL catalog, as illustrated in the right panels of figure~\ref{fig:dla_bal_splits}. The baseline measurement (BASE) uses a catalog in which quasars tagged as BAL with the $BI > 0$ criterion have been removed. The upper panel of the figure compares it to an analysis variation (AICUT) with the $AI > 0$ cut, which removes $22$ \% of quasars, including the $4$ \% of $BI > 0$ quasars. We observe a near $5$ \% difference at small scales, well above the level of statistical uncertainties, giving a $0.0$ Correlation $p$-value. However, removing all $AI$ BAL quasars would induce a large drop in statistics, and we aim to match the power spectrum obtained with this severe cut while keeping most of the quasars. 

Two additional variations are then performed: masking at the spectrum level all $AI$ BAL absorptions while keeping all quasars in the catalog (AIMASK) and masking all $AI$ BAL pixels while removing $BI$ BAL quasars from the catalog (BIC\_AIM). AIMASK is a drastic change relative to baseline as no quasars are removed from the input catalog, while BIC\_AIM can be considered as the baseline while additionally masking for BAL regions that are passing $AI > 0$ criterion. As shown on the lower two right panels of the same figure, the impact of masking BAL is very similar for AIMASK and BIC\_AIM: it decreases the $\pk$ level for all scales due to the effect of pixel masking while accounting for the small scale difference due to the unaccounted $AI$ BAL that are masked. Since the impact of masking is way more prominent when the $BI$ BAL quasars are kept (AIMASK), we choose BIC\_AIM as our new baseline.

To verify that the impact of $AI$ BAL quasars is correctly accounted for, we compare our new baseline to the case where all $AI$ BAL quasars are removed from the catalog (AICUT). The new baseline (NEWBASE) is corrected for the impact of BAL masking using the correction derived in \validpaper. The number of $AI$ BAL quasars is larger in the \drone~data than in the mocks used in \validpaper. We normalize the absolute level of the BAL masking correction to account for the larger fraction of BAL in the data. The resulting new baseline (BIC\_AIM with corrections) is compared to the $AI$ cutting case in figure~\ref{fig:new_bal_split}. Compared to the first left panel of figure~\ref{fig:dla_bal_splits}, the new baseline properly accounts for the $AI$ BAL, as shown by the $p=0.05$ value. This p-value is the lower value for six bins in redshift, so $p=0.05$ indicates that the test is passed for all redshift bins. We will use this study of the impact of $AI$ BAL to derive systematic uncertainties linked to BAL completeness and the BAL masking correction.

\subsection{FFT method parameters variation}
\label{subsec:method_split}

\begin{figure}
    \centering
    \includegraphics[width=0.49\columnwidth]{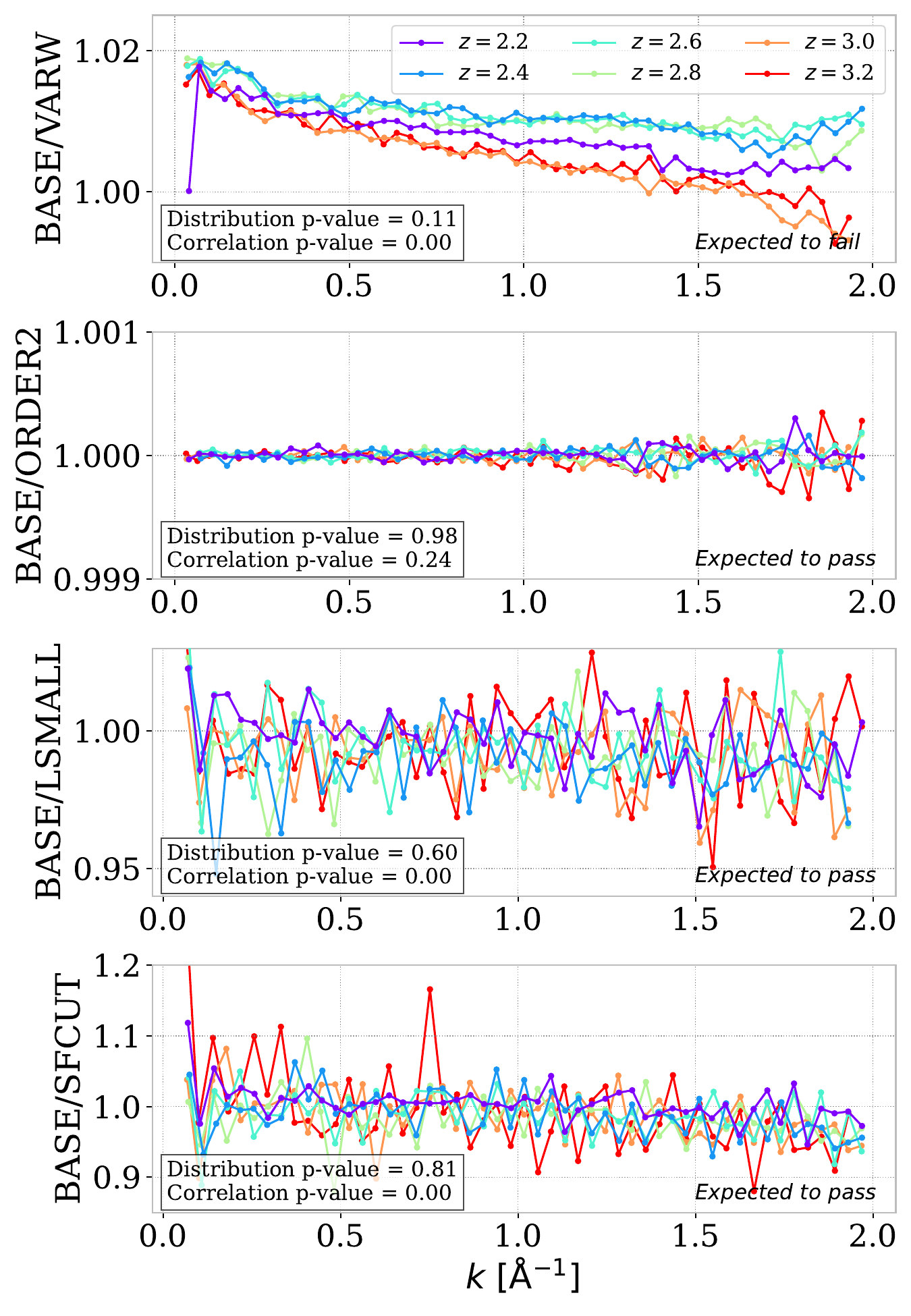}
    \includegraphics[width=0.49\columnwidth]{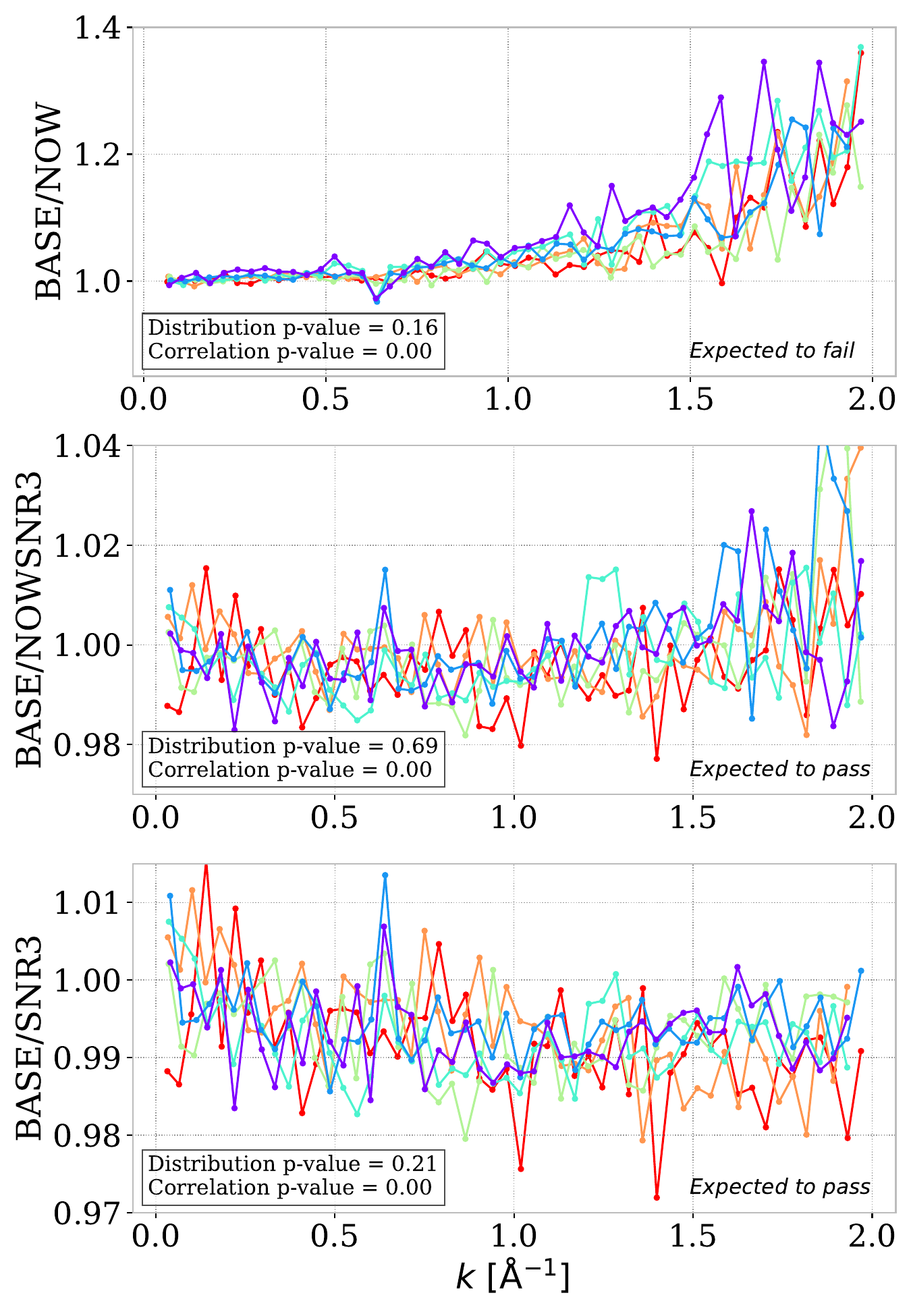}
    \caption{Ratios of the baseline power spectrum (BASE) to the power spectrum obtained from analysis variations. (left) Variation of the continuum fitting method: using wavelength-dependent correction terms in the computation of the continuum fitting weights (VARW), using a second-order polynomial to compute the quasar-specific continuum (ORDER2), taking a shorter definition of the \lya~forests in terms of rest-frame wavelength (LSMALL), and performing the continuum fitting independently on the different sub-forests with a zero-order polynomial for the quasar-specific continuum (SFCUT). (right) Variation of the $\snr$ cut and the $\pk$ averaging method: Performing the average without weights and a $\snr > 1$ cut (NOW), without weights and an $\snr > 3$ cut (NOWSNR3), and with the $\snr$ weighting scheme used for baseline and a $\snr > 3$ cut (SNR3).}
    \label{fig:method_splits}
\end{figure}

We vary some parameters chosen as fixed in section~\ref{sec:method} to blindly search for potential sources of analysis systematic error. First, the parameters used for continuum fitting in section~\ref{subsec:continuum} are varied, and the ratio with baseline is shown in the left panels of figure~\ref{fig:method_splits}. We carry out the continuum fitting while allowing noise-dependent weights in the likelihood used for fitting (VARW). In equation~\ref{eq:likelihood}, the weights $\sigma_q = \overline{F}(\lambda_i) C_{\mathrm{q}}\left(\lambda_{i},z_{\mathrm{q}}, a_{\mathrm{q}}, b_{\mathrm{q}}\right)$ in the likelihood are replaced by:

\begin{equation}
\begin{split}
\frac{\sigma_{\mathrm{q}}^2}{\left(\overline{F}(\lambda_i) C_{\mathrm{q}}\left(\lambda_{i},z_{\mathrm{q}}, a_{\mathrm{q}}, b_{\mathrm{q}}\right)\right)^2} &=\eta(\lambda) \frac{\sigma_{\mathrm{pip}, \mathrm{q}}^2(\lambda)}{\left(\overline{F}(\lambda_i) C_{\mathrm{q}}\left(\lambda_{i},z_{\mathrm{q}}, a_{\mathrm{q}}, b_{\mathrm{q}}\right)\right)^2}+\sigma_{\mathrm{lss}}^2(\lambda) \\ & + \epsilon(\lambda) \frac{\left(\overline{F}(\lambda_i) C_{\mathrm{q}}\left(\lambda_{i},z_{\mathrm{q}}, a_{\mathrm{q}}, b_{\mathrm{q}}\right)\right)^2}{\sigma_{\mathrm{pip}, \mathrm{q}}^2(\lambda)}\, ,
\end{split}
\end{equation}

\noindent with additional wavelength-dependent parameters $\eta$, $\sigma_{\mathrm{lss}}$ and $\epsilon$, as it is the case for BAO analysis. In a second variation, we added a $c_q \lambda^2$ term to compute individual quasar continuum in equation~\ref{eq:continuum_quasar} (ORDER2). In a third variation we modify the minimum and maximum wavelength considered in the computation of flux contrast from $1,050 < \lambda_{\mathrm{rf}} < 1,180$ \AA~to the reduced range $1,070 < \lambda_{\mathrm{rf}} < 1,160$ \AA~(LSMALL). Finally, we perform the sub-forest cut before the continuum fitting and fit the continuum independently for each sub-forest (SFCUT). For the latter variation, in comparison to baseline, we also remove the $b_q \lambda$ in equation~\ref{eq:continuum_quasar} as this term would generate a broken linear continuum fitting when separated in sub-forests and would suppress small-wavenumber modes. Note that the SFCUT variation corresponds to how continuum fitting was performed for eBOSS~\cite{chabanier_one-dimensional_2019}.

The four variations considered do not induce a significant decrease in the number of sub-forests: only a nearly $5$ \% decrease for LSMALL due to the creation of shorter sub-forests that do not pass the size cuts. The second test (ORDER2) has a very small effect on $\pk$ as indicated by the reported $p$-values. All the three other tests (VARW, LSMALL, SFCUT) shows some statistically significant correlation between the ratio and the wavenumber. The VARW test has a small but significant effect on the power spectrum. However, including $\sigma^2_{\rm LSS}$ in the weights is biasing the evaluation of the power spectrum, so this effect is not surprising. The LSMALL and SFCUT variations show more significant fluctuations. We interpret those spikes as caused by adding smaller sub-forests that appear as outliers in the averaging, and that are potentially not accounted by the outlier finder detailed in section~\ref{sec:data}. Those two variations should be considered in a cosmological interpretation study as they could potentially yield different results.

In a separate study, we test the average $\pk$ computation by varying the $\snr$ cut and the weighting scheme. The right panels of figure~\ref{fig:method_splits} show the variation performed here: using a constant weighting scheme with the same $\snr > 1$ cut (NOW), with a $\snr > 3$ cut (NOWSNR3) which is close to what was done in eBOSS~\cite{chabanier_one-dimensional_2019}, or using the same $\snr$ weighting scheme as baseline with the $\snr > 3$ cut (SNR3). The $\snr > 3$ cut implies keeping only between $32$ to $40$ \% of the \drone\ sub-forests depending on the redshift bin. It corresponds to a drastic cut in statistics, but can be considered a robust sample for small scales, for which noisy sub-forests can significantly impact the measurement. Indeed, noisy sub-forests have a large noise power spectrum and a slight noise misestimation causes a large small-scale variation. 
The upper plot exhibits a large discrepancy between the baseline and the case without weighting. 

For the second plot when a $\snr > 3$ cut is applied to the unweighted analysis, the "Distribution" p-value is improved, but there still exists a wavenumber correlation. However, we note that the ratio is very close to unity with a maximal value near $2$ \%. The NOW discrepancy can therefore be ascribed to noisy spectra with $1<\snr<3$, and probably mostly to spectra with $\snr\gtrsim 1$. These spectra have a very small weight and a negligible contribution to the baseline power spectrum, which, therefore, is hardly affected by removing $\snr<3$ spectra, as illustrated by the lower plot, which shows a small (near $1$ \%) difference, which is well below the level of small-scale systematics (e.g., resolution) associated to the final measurement in section~\ref{subsec:syst}. Even if the impact of biasing caused by low $\snr$ sub-forest is mostly accounted for by our $\snr$-weighting, the $\snr>3$ can be considered a potential variation for cosmological interpretations.

\subsection{Comparison of exposure numbers}
\label{subsec:exp_split}

\begin{figure}
    \centering
    \includegraphics[width=0.49\columnwidth]{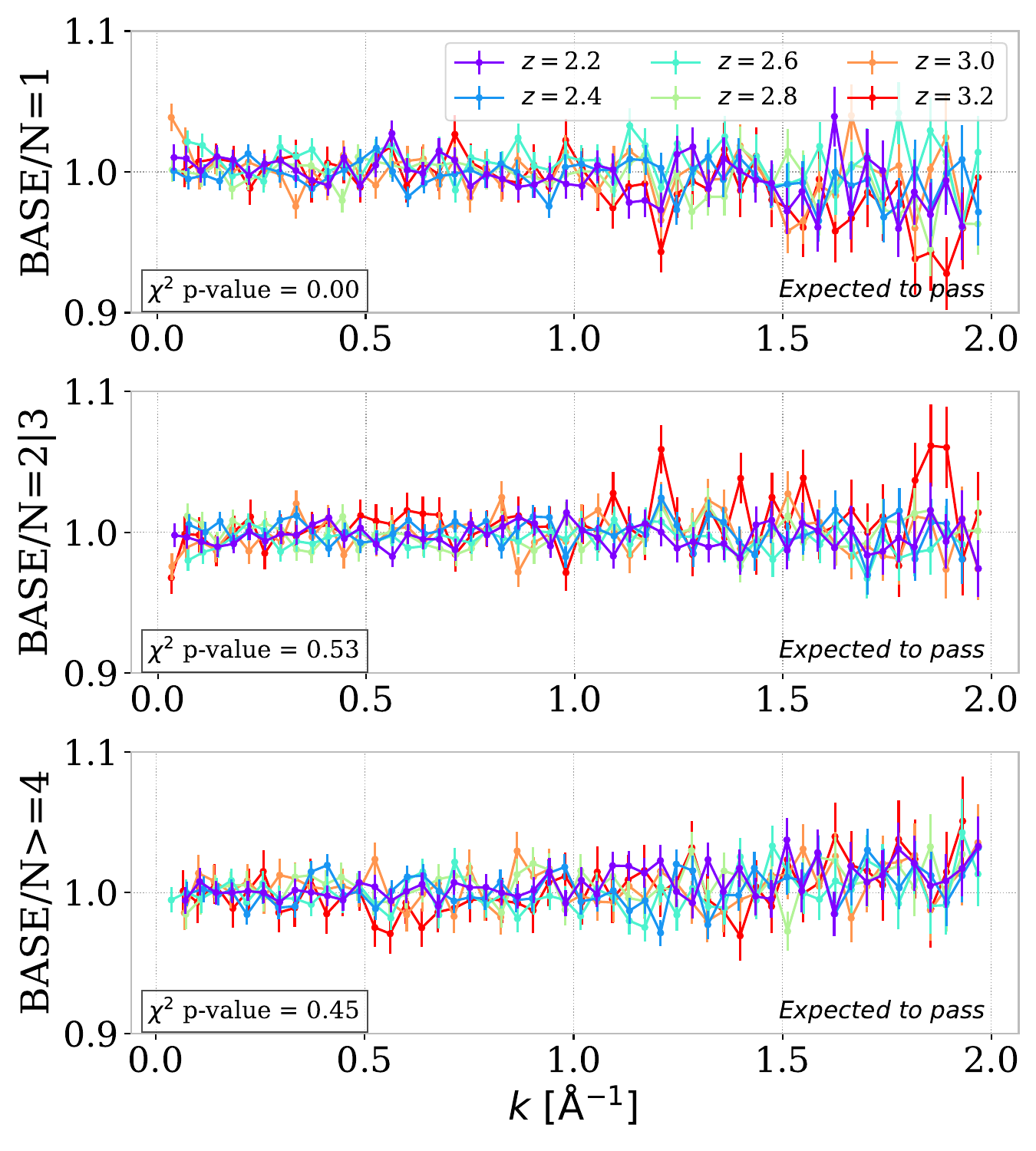}
    \includegraphics[width=0.49\columnwidth]{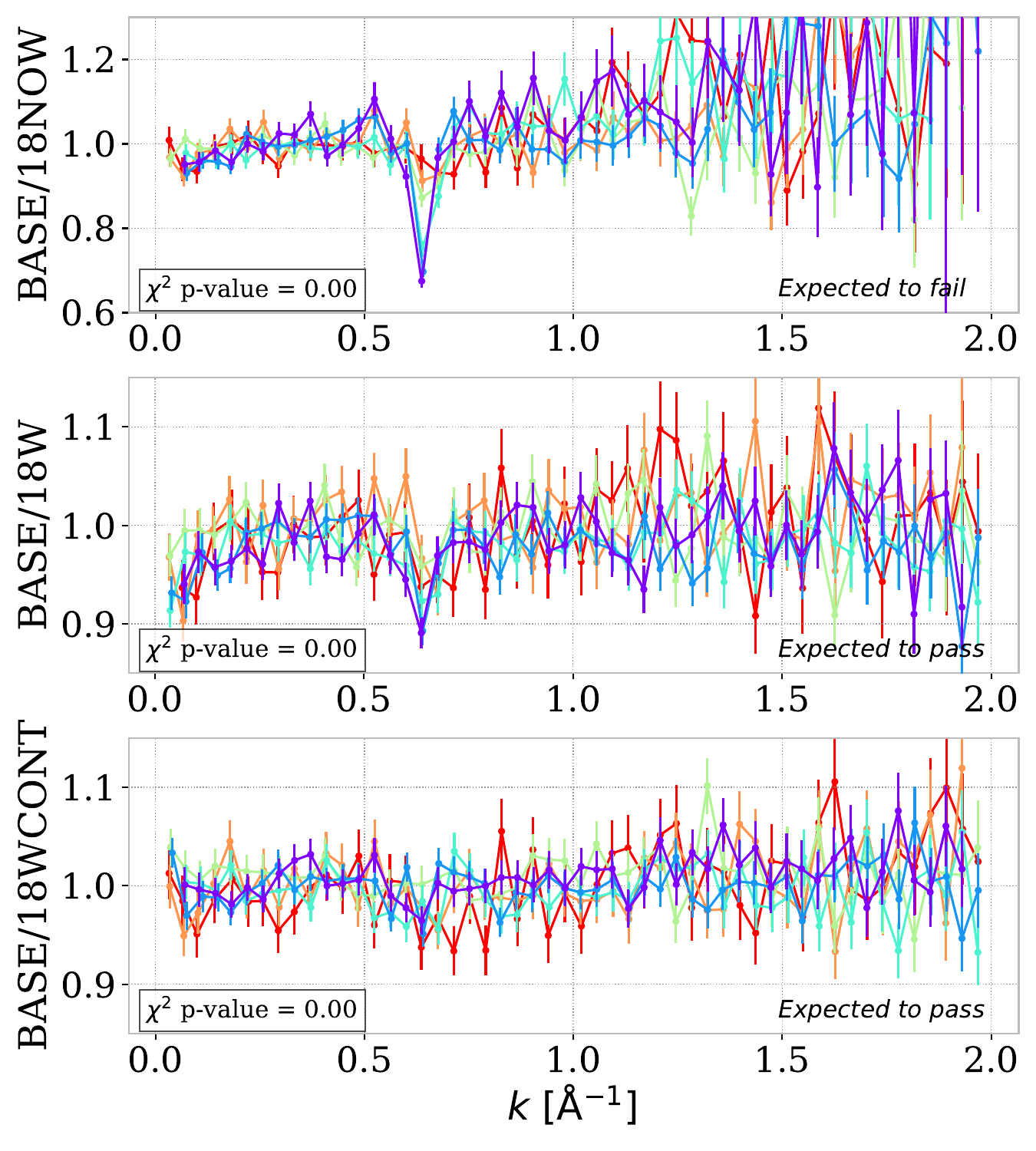}
    \caption{Ratios of the baseline power spectrum (BASE) to the power spectrum obtained from analysis variations. (left) Splitting of the quasar catalog between quasars observed with one (N=1), two or three (N=2$|$3), and four or more (N$>$=4) exposures. (right) Splitting of the amplifier sets used to compute $\pk$. We show the ratio for the power spectrum measured using the $18^{\mathrm{th}}$ amplifier set, which exhibits a noticeable discrepancy with the baseline. A variation of weighting scheme and continuum fitting are performed: without averaging weights (18NOW), with $\snr$ weighting (18W), and with $\snr$ weighting but taking the continuum fitting measure on the baseline for the $18^{\mathrm{th}}$ amplifier set power spectrum (18WCONT). The amplifier-related failures in the column are tackled in the final analysis by using the cross-exposure estimator for $z \leq 2.7$.}
    \label{fig:nexp_amp_splits}
\end{figure}

Each quasar is planned to be observed several times by the DESI instrument, and since \drone~is an intermediate data set, it contains spectra observed with very different numbers of exposures. The coadding of exposures can potentially introduce a bias, especially when it involves a large number of exposures with very different integration times and observing conditions. A data split involving quasars observed over only one (N=1), two or three (N=2$|$3), or four or more (N$>$=4) exposures is shown in the left panels of figure~\ref{fig:nexp_amp_splits}. The data splits contain, on average over redshift bins, $39$ \%, $36$ \%, and $25$ \% of the total number of sub-forests, respectively, thus resulting in larger error bars. The ratios do not show strong variations from unity; however, only the (N=2 $|$3) and (N $>$=4) are passing the $ \chi^2 $ test. The low number of exposures (N=1) can potentially bias our measurement, especially at small scales, because a quasar with a low number of exposures tends to be noisier, thus giving different weights when averaging. Removing this sample implies a too-large decrease in statistics, but this variation should be considered for cosmological interpretation and future measurement with larger data sets and a larger number of exposures.

\subsection{Measurement with different amplifier sets}
\label{subsec:amp_split}

The focal plane of the DESI instrument is composed of ten modules called petals, each containing 500 robotic fiber positioners. The light measured by those positioners is linked to ten spectrographs by optical fibers. Each spectrograph receives the fiber associated with one petal and scatters the received light using volume phase holographic gratings. Three CCD cameras per spectrograph read the spectra, covering different wavelength ranges. Finally, four amplifiers per CCD are positioned in squares to perform the readout. 
In the $x$ direction, a CCD image contains different spectra chunks, while $y$ is the direction of wavelengths. 
A given quasar spectrum is then measured over six amplifiers (2 per CCD camera). For one spectrograph, the fibers directed to the left and right parts of the CCD cameras constitute two sets, measured with completely distinct amplifiers. At the spectrum level, the data can be split between 20 amplifier sets, according to the left and right parts of the ten spectrographs. 

We measured $\pk$ over those 20 amplifier set splits independently, which constitutes a significant drop in terms of statistics. The right panels of figure~\ref{fig:nexp_amp_splits} show the ratio between baseline and the power spectrum using only spectra from the amplifier set number 18 with constant-weight averaging (18NOW), with $\snr$-weight averaging (18W), and with the $\snr$-weight averaging but using the continuum from the complete \drone~data set (18WCONT). The 18$^{\mathrm{th}}$ amplifier set was chosen because it shows a considerable difference for low redshift near wavenumber $k \sim 0.65$ \invAA. Among the 20 splits of the amplifier set, only two exhibit this behavior, with the 16$^{\mathrm{th}}$ data split showing a smaller spike amplitude. Those spikes originate from oscillatory patterns of period $\Delta \lambda = 9.6$ \AA (corresponding to $k \sim 0.65$ \invAA)~present on \drone~calibrations images used for bias corrections.

The source of those patterns was not identified and was highly complicated to correct due to their partial coverage of the amplifier and sporadic occurrence. In contrast to the companion paper \qmleyone, the impact of this feature drastically decreases when using the $\snr$ weighting scheme. Furthermore, comparing 18W and 18WCONT indicates that a subsequent part of the spike is caused by the continuum fitting procedure itself when using only the 18$^{\mathrm{th}}$ amplifier set. The spike is more prominent at small redshifts. As pointed out in section~\ref{subsec:noise}, the cross-exposure estimator accounts for this spike because it uses the individual cross-power spectrum between different fibers and, consequently, different amplifier sets. As we use the cross-exposure estimator for the lower redshift ($z < 2.7$) where the $0.65$ \invAA~spike is the largest ($2.3\sigma$ for $z=2.2$ and $3.5\sigma$ for $z=2.4$ and dropping to $1\sigma$ or below for larger redshifts), we consider that this feature does not impact our measurement, and we choose not to add a dedicated systematic uncertainty. We interpret the low $p$-value for the last panel to be mostly caused by fluctuations and not by the $0.65$ \invAA~spike itself.

\subsection{Galactic hemisphere split}
\label{subsec:gc_split}

\begin{figure}
    \centering
    \includegraphics[width=0.49\columnwidth]{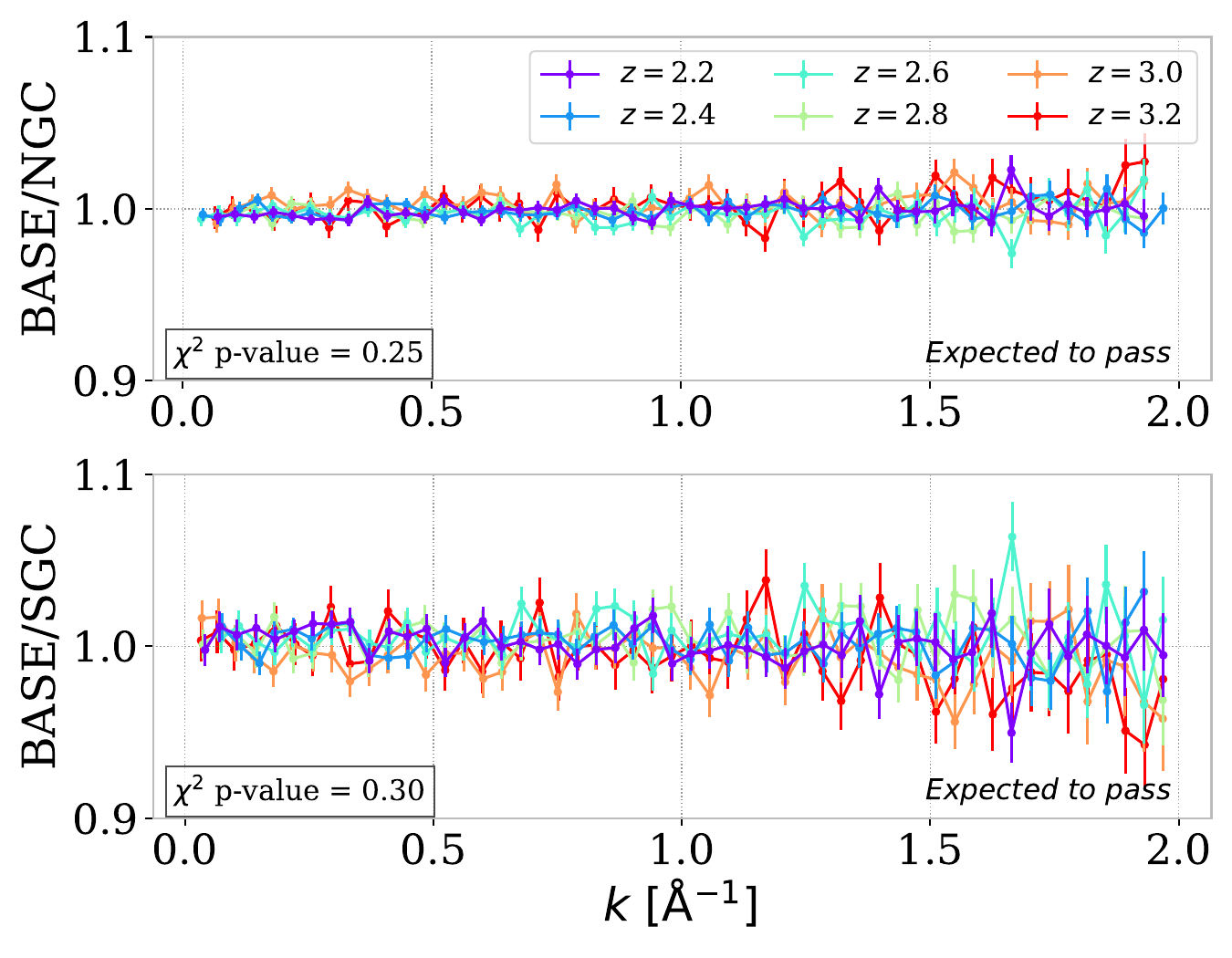}
    \includegraphics[width=0.49\columnwidth]{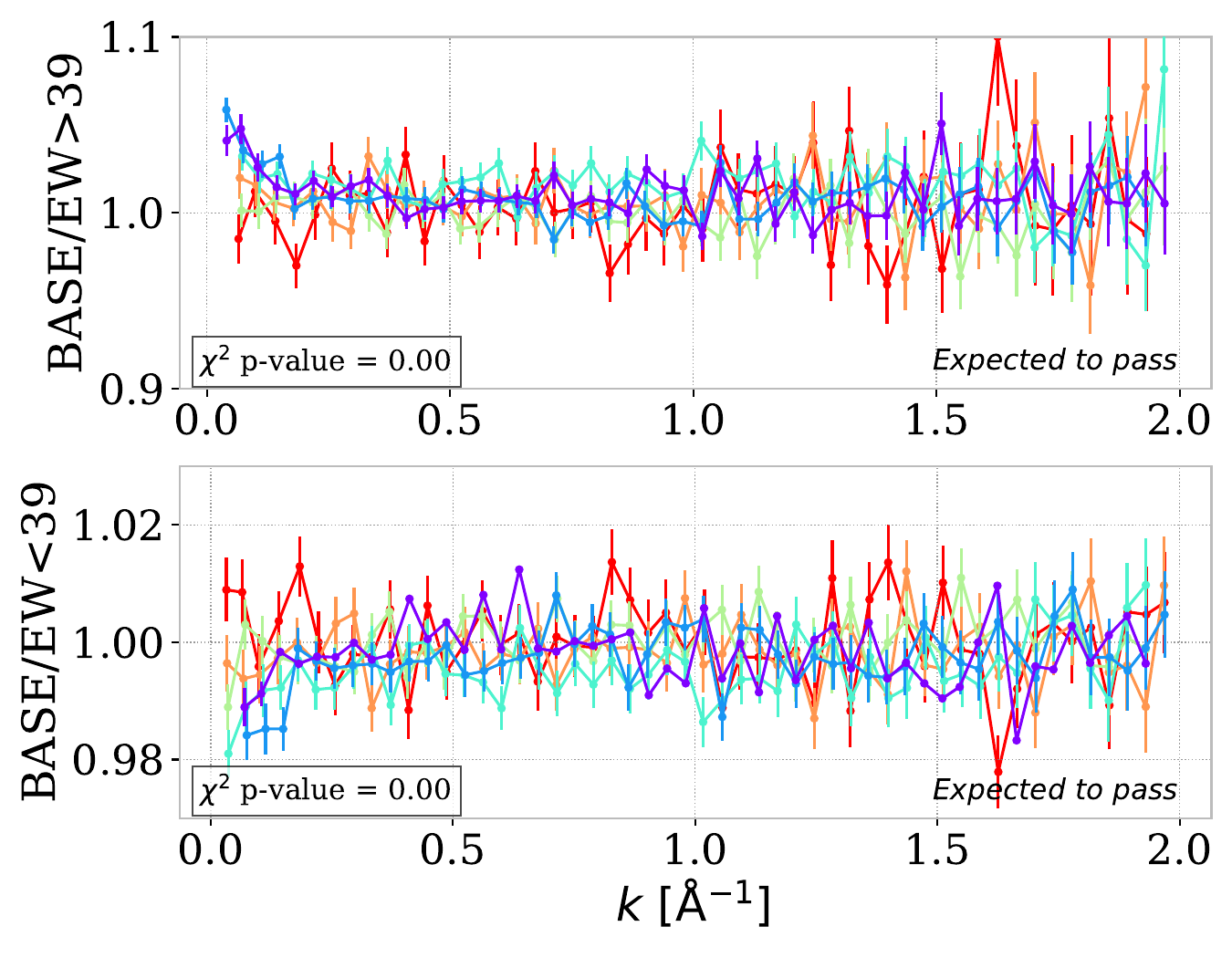}
    \caption{Ratios of the baseline power spectrum (BASE) to the power spectrum obtained from analysis variations. (left) Splitting of the quasar catalog between north (NGC) and south (SGC) galactic cap. (right) Splitting of the quasar catalog based on the \civ~equivalent width noted EW. The split is performed such that the quasar (but not the sub-forest) catalog is split in half: EW$>$39 and EW$<$39.}
    \label{fig:gc_ew_splits}
\end{figure}

Different imaging surveys are used for the quasar target selection of the DESI survey~\cite{dey_overview_2019,yeche_preliminary_2020,chaussidon_target_2022}. Inhomogeneities in the photometric target pre-selection can significantly impact galaxy clustering studies, since they affect the target density and subsequently the estimation of the matter density. We do not expect an impact on the measurement of $\pk$ since each spectrum pixel is a proxy for the IGM density independently of the quasar density. However, as a sanity check, we still perform a data split between the North Galactic Cap (NGC) and the South Galactic Cap (SGC), as shown on the left panels of figure~\ref{fig:gc_ew_splits}. The NGC dataset contains $65$ \% of the sub-forests, and the SGC $35$ \%. As expected, the $\pk$ measured on both data splits show no visible difference with the baseline, as also indicated by the reported $p$-values.

\subsection{CIV equivalent width}
\label{subsec:ew_split}

The last data split performed follows the BAO analysis~\cite{DESIBAOlya2024,DESIBAOlya2025} and is related to the \civ~equivalent width (EW) in angstrom, i.e.\ the spectral area associated with the quasar \civ~emission line. The EW variable is a good quasar population discriminant because of the Baldwin effect, which causes EW to be anti-correlated to the quasar continuum luminosity~\cite{Baldwin1977}. Consequently, the high EW quasars will tend to be noisier. We use the EW measurement performed with \texttt{fastspecfit}\git{desihub/fastspecfit}{} in~\cite{DESIBAOlya2024,DESIBAOlya2025} to create two populations equal in terms of quasar number: EW$>$39 and EW$<$39. The result of $\pk$ computations for those two EW populations is shown in the right panels of figure~\ref{fig:gc_ew_splits}. Since we apply a $\snr > 1$ quality cut, the noisier EW$>$39 measurement contains only $35$ \% of the total sub-forests. The EW$>$39 data set shows a $5$ \% discrepancy for small redshift and $k < 0.2$ \invAA. This discrepancy can be interpreted as a proximity effect impacting the larger scales of the quasar spectra. The EW$<$39 sample is also showing a discrepancy, shown by the $p$-value, but with a smaller level (near $2$ \% at large scales). Removing the EW$>$39 sample would imply a statistical drop that is too large. For this analysis, we keep the two populations of quasars in the $\pk$ baseline measurement. However, the EW$<$39 data sample should also be considered as a variation for cosmological analysis, and this EW test should be investigated more in future studies. For now, the level of the $\sim 1 \%$ induced bias compared to EW$<$39 is well below the level of large-scale systematics such as DLA completeness detailed in section~\ref{subsec:syst}.

\subsection{Analysis variation and data split conclusion}
\label{subsec:split_conclusion}

\begin{table}
    \centering
    \begin{tabular}{c|c|c | c | c | c | c|c}
        \hline
        Variation & {\scriptsize NEWBASE} & {\scriptsize LSMALL} & {\scriptsize SFCUT} & {\scriptsize SNR $>$ 3} & {\scriptsize (N=2$|$3)} & {\scriptsize (N$>$=4)} & {\scriptsize EW $<$ 39} \\
        \hline
        \hline
        Sub-forests & 99887 & 96526 & 80261 & 35624 & 36155 & 26328 & 67792 \\
        \hline
        Ratio $\pk$ $k_{\mathrm{low}}$ & 0.977 & 0.966 & 0.942 & 1.002 & 0.998 & 1.002 & 1.005 \\
        Ratio $\pk$ $k_{\mathrm{med}}$ & 0.992 & 0.990 & 0.983 & 1.007 & 0.998 & 1.001 & 1.001 \\
        Ratio $\pk$ $k_{\mathrm{high}}$ & 0.993 & 1.017 & 1.035 & 1.005 & 1.003 & 0.987 & 1.001 \\
        \hline
        Ratio $\sigma_{1\mathrm{D},\alpha}$ $k_{\mathrm{low}}$ & 0.986 & 1.651 & 2.822 & 1.314 & 1.650 & 1.859 & 1.176 \\
        Ratio $\sigma_{1\mathrm{D},\alpha}$ $k_{\mathrm{med}}$ & 1.000 & 1.246 & 1.690 & 1.142 & 1.649 & 1.773 & 1.134 \\
        Ratio $\sigma_{1\mathrm{D},\alpha}$ $k_{\mathrm{high}}$ & 0.996 & 1.280  & 1.727 & 0.960 & 1.641 & 1.622 & 1.099 \\
        \hline
    \end{tabular}
    \caption{Number of sub-forests, ratio of the one-dimensional power spectrum, and ratio of statistical error bars between the analysis variation in comparison to the baseline (BASE). Those ratios are expressed for a specific redshift bin ($z=2.4$) as the ratio of averages on a specific wavenumber range. The three different wavenumber ranges used are $k_{\mathrm{low}} = [0.0, 0.5]$ \AA$^{-1}$evaluated, $k_{\mathrm{med}} = [0.75, 1.25]$ \AA$^{-1}$, and $k_{\mathrm{med}} = [1.5, 2.0]$ \AA$^{-1}$. For reference on the same redshift bin, the baseline (BASE) number of sub-forests is $100357$, average $\pk$ values are $0.306$, $0.143$, and $0.078$ \AA~for the $k_{\mathrm{low}}$, $k_{\mathrm{med}}$, and $k_{\mathrm{high}}$ wavenumber ranges respectively. The statistical error bars for the same wavenumber ranges are $0.0017$, $0.00096$, and $0.0010$ \AA.}
    \label{tab:ccl_analysis}
\end{table}

We performed an extensive data split study to find potential biases in our $\pk$ measurement. Those data splits allow us to validate our measurement and improve the characterization of systematic uncertainties. Notably, this study indicates that the impact of $AI$ BAL on the $\pk$ measurement cannot be ignored, and we changed the baseline measurement to mask the associated absorptions, and correct for the masking (NEWBASE). This analysis shows that some effects are slightly changing the global trend of $\pk$, and some imprint fluctuations sufficient not to pass our statistical tests. We can also note that the "Correlation $p$-value" test is especially severe. Removing the sub-samples that do not pass the tests would imply a significant statistical drop; we then keep them for the new baseline used in this paper (NEWBASE). First, those effects (continuum fitting variations, $\snr$ cut, number of exposures, and \civ~EW quasar population) should be considered as baseline variations for any cosmological interpretation. Secondly, those sub-samples should potentially be removed for future DESI releases with even larger data sets available.

To give a more quantitative view of the analysis, we provide in table~\ref{tab:ccl_analysis} the number of sub-forests, the ratios of $\pk$, and statistical error for the variations that we considered relevant as alternatives for cosmological interpretations. Those properties are expressed for a redshift bin $z=2.4$ in low, medium, and high wavenumber ranges. The resulting power spectrum for all those variations is also provided in the data release associated with this paper. The NEWBASE version gives very similar ratios except for $\pk$ at low wavenumber, which is expected after changing the treatment of BAL quasars. Changing the continuum fitting method (LSMALL and SFCUT) considerably increases the statistical error bar and can impact the $\pk$ slope and thus $n_{\rm s}$ as highlighted by the change in $\pk$ ratios. Finally, the four data splits (SNR$>$3, (N=2$|$3), (N$>$=4), and EW $<$ 39) do not drastically change the $\pk$ ratios over different redshift ranges, and should have a negligible impact on cosmological interpretation. We note that for the selection of higher quality spectra (SNR$>$3 and EW $<$ 39), the statistical error bar is very similar, despite the use of way fewer sub-forests. The detailed study of the impact of the analysis variation on cosmological parameters is out of scope and will be treated in an upcoming paper.

\section{Power spectrum uncertainties}
\label{sec:uncertainties}

\subsection{Statistical uncertainties and covariance matrix}
\label{subsec:stat}

\begin{figure}
    \centering
    \includegraphics[width=\linewidth]{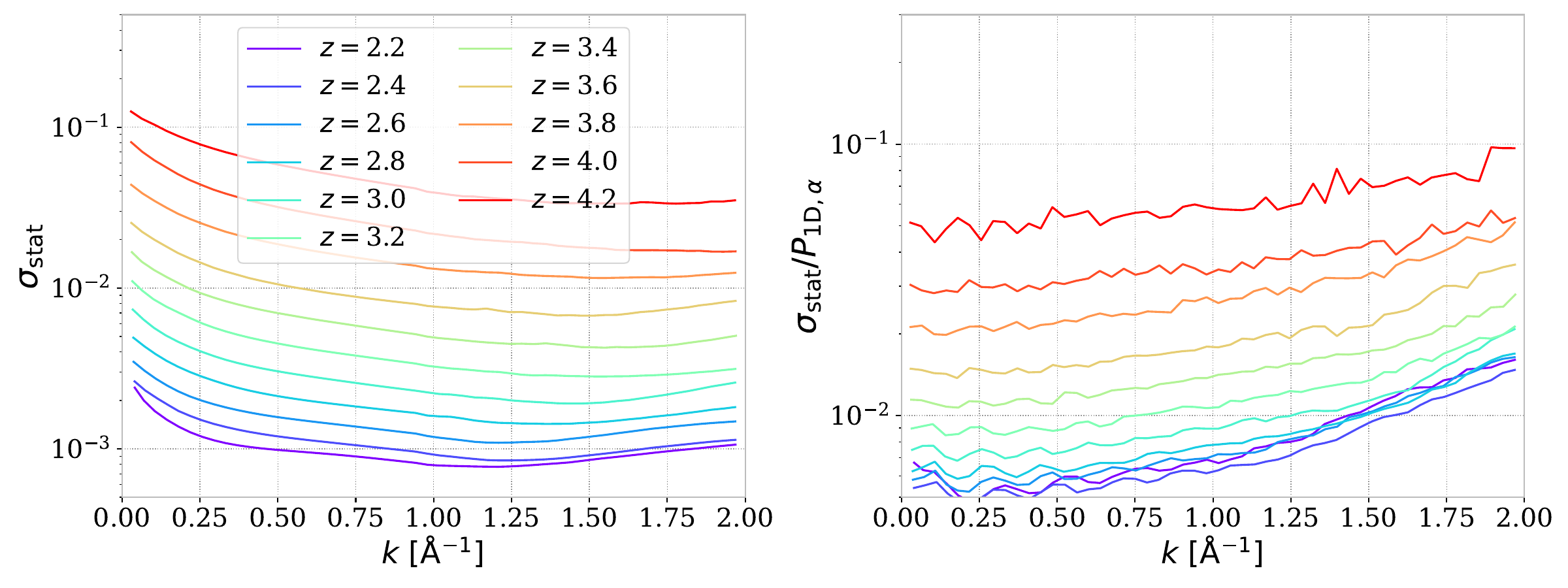}
    \caption{Statistical uncertainties of the \drone~$\pk$ measurement, computed using equation~\ref{eq:var}. The uncertainties are smoothed in each redshift bin using a Savitzky-Golay filter. The left panel shows the absolute level of those uncertainties, and the right panel shows their level relative to the measured power spectrum.}
    \label{fig:stat_error}
\end{figure}

As the aim of this \drone~measurement is to be interpreted in terms of cosmological parameters, we estimate the covariance matrix associated with $\pk$. Furthermore, we improve the study conducted in~\desiedrfft~concerning the estimation of statistical and systematic uncertainties. The full derivation of the variance and covariance estimators used here is presented in Appendix~\ref{appendix:cov}.

The expression of the power spectrum $\pk(A,z)$ involves a weighted average of $\ps(k)$ (Eq.~\ref{eq:p1d_estimator}). In Appendix~\ref{appendix:cov}, we derive an estimator for the variance of a weighted average $\sum w_i A_i  / \sum w_i$  and for the covariance between two such weighted averages. However, this derivation requires that the $A_i$ be uncorrelated. When a sub-forest $s$ has several wavenumbers $k$ in a given $A$ bin, these wavenumbers are correlated, so we cannot use our variance estimator for the weighted average $\langle \ps(k) \rangle_{s\in z, k \in A}$.

We therefore introduce $A_s$ as the ensemble of wavenumbers associated with sub-forest $s$ that fall into bin $A$, $N_{A,s}$ the number of wavenumbers in $A_s$, and $P_{A,s} = \sum_{k \in A_s} \ps(k) / N_{A,s}$ the average of individual power spectra in bin $A$. We can neglect the correlation between different sub-forests as accounting for them would give a three-dimensional estimator. Therefore, we use our variance estimator for the $P_{A,s}$. We introduce the weights  $w_{A,s}^\prime = N_{A,s} w_{A,s}$ to rewrite  $\langle \ps(k) \rangle_{s\in z, k \in A}$ as a weighted average of the $P_{A,s}$:

\begin{equation}
\langle \ps(k) \rangle_{k\in A,s\in z} 
= \frac{\sum_{k\in A,s\in z}w_{A,s} \ps(k)}{\sum_{k\in A,s\in z} w_{A,s}}
= \frac{\sum_{s\in z}w_{A,s}^\prime P_{A,s}} {\sum_{s\in z} w_{A,s}^\prime} 
= \left\langle P_{A,s} \right\rangle_{s \in z}\, .
\end{equation}

We can then use our estimator for the variance 

\begin{equation}
\label{eq:var}
\sigma_{1\mathrm{D},\alpha}^2 (A,z) =  \left(\frac{(\sum_{s \in z} w'_{A,s})^2 }{\sum_{s \in z} (w'_{A,s})^2}  -1\right)^{-1} \left[ \frac{\sum_{s \in z} (w'_{A,s})^2 P_{A,s}^2}{\sum_{s \in z} (w'_{A,s})^2} - \left\langle P_{A,s} \right\rangle_{s \in z}^2 \right]\,,
\end{equation} 

\noindent and the covariance

\begin{align}
\label{eq:cov}
\mathcal{C}_{1\mathrm{D}, \alpha}(A, B, z) = & \left(\frac{\sum_{s \in z} w'_{A,s} \sum_{s \in z} w'_{B,s}}{\sum_{s \in z} w'_{A,s} w'_{B,s}}  -1\right)^{-1} \nonumber \\
&\left[ \frac{\sum_{s \in z} w'_{A,s} w'_{B,s} P_{A,s} P_{B,s}}{\sum_{s \in z} w'_{A,s} w'_{B,s}} - \left\langle P_{A,s} \right\rangle_{s \in z} \left\langle P_{B,s} \right\rangle_{s \in z}\right]\,.
\end{align}

We note that our estimator fluctuates significantly and can even result in a negative variance for a small dataset size (at high redshift in our case). This issue is solved by smoothing the variance with a Savitzky-Golay filter implemented in \texttt{scipy}~\cite{SciPy}. Similarly, the covariance matrix is smoothed along its two dimensions using a Savitzky-Golay filter implemented in \texttt{sgolay2} \git{espdev/sgolay2}{}. The covariance and variance estimators have been validated in the companion \validpaper~paper by comparing the variation between $20$ \drone-like uncontaminated mocks with the average of those mocks. This validation showed that our estimator slightly overestimates the covariance by $10$ \%. We correct our covariance matrix by renormalizing by this factor, independently of redshift.

Figure~\ref{fig:stat_error} shows the standard deviation calculated with equation~\ref{eq:var} for the \drone~power spectrum. Compared to \desiedrfft~results, the uncertainties are smoother due to the applied smoothing but have a similar profile. The level of uncertainties is much lower than for \desiedrfft: between $2.6$ and $3$ times decrease depending on redshift. This improvement is even higher than the five times increase in quasar statistics, which should decrease the error bar only by $\sim 2.23$ times. Considering that $\snr$ weighting is used in both measurements, we interpret this significant improvement as caused by decreasing noise in the input data set.

\begin{figure}
    \centering
    \includegraphics[width=\linewidth]{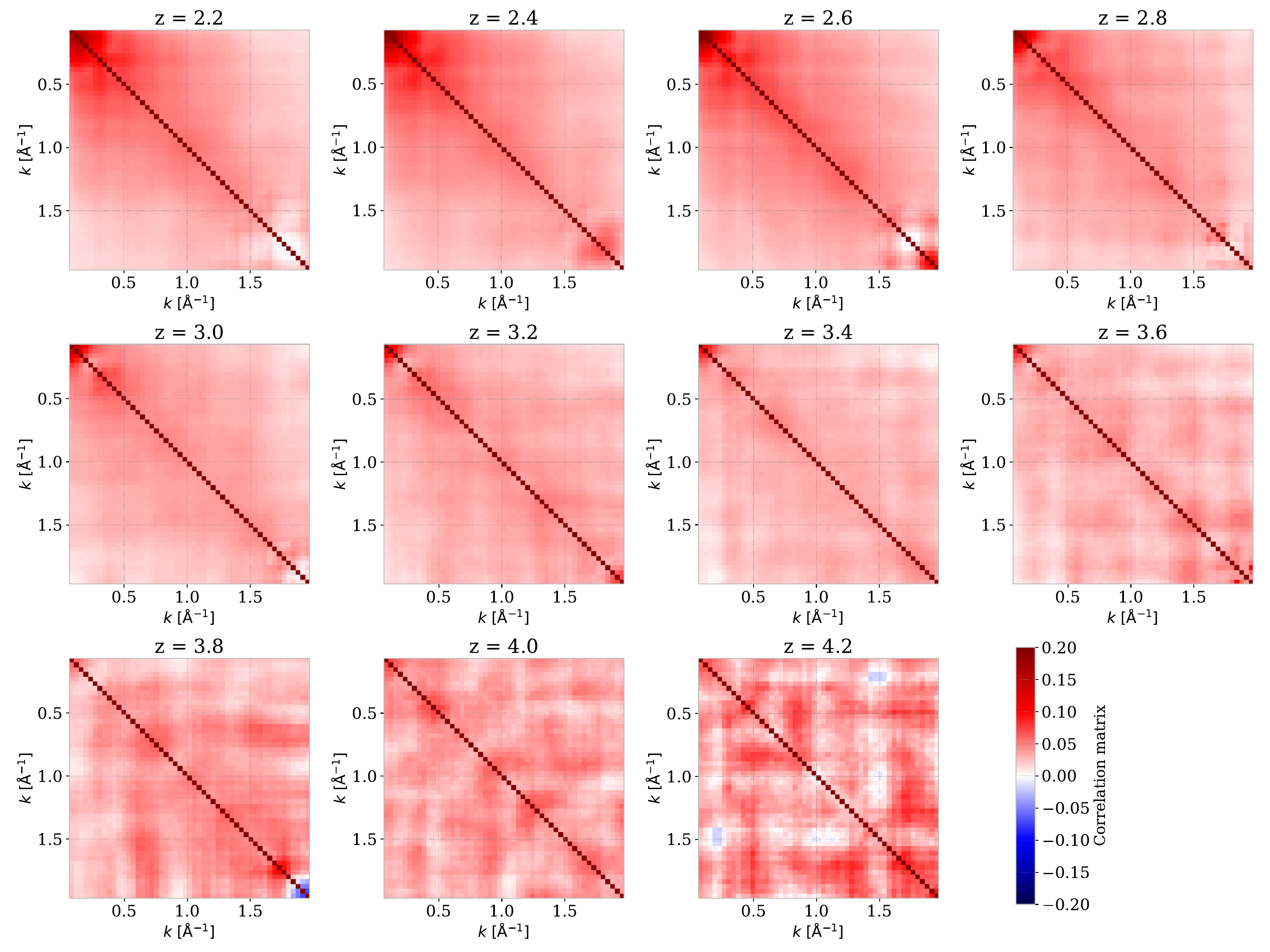}
    \caption{Correlation matrix of the statistical error of the \drone~power spectrum measurement, obtained by normalizing the covariance in equation~\ref{eq:cov}. The covariance is computed for the new $\pk$ baseline and smoothed with a 2D Savitzky-Golay filter.}
    \label{fig:covariance}
\end{figure}

Figure~\ref{fig:covariance} shows the correlation matrix for the \drone~data set. The correlation profile is very similar to the one obtained for eBOSS in \cite{chabanier_one-dimensional_2019}, with a significant correlation (near $20$ \%) at small wavenumber and decreasing as a function of redshift. There are some instabilities for high redshift caused by the low number of available sub-forests. The cosmological interpretation of this measurement should account for those instabilities.

\subsection{Systematic uncertainties}
\label{subsec:syst}

\begin{figure}
    \centering
    \includegraphics[trim=0cm 3cm 0cm 0cm, width=0.9\linewidth]{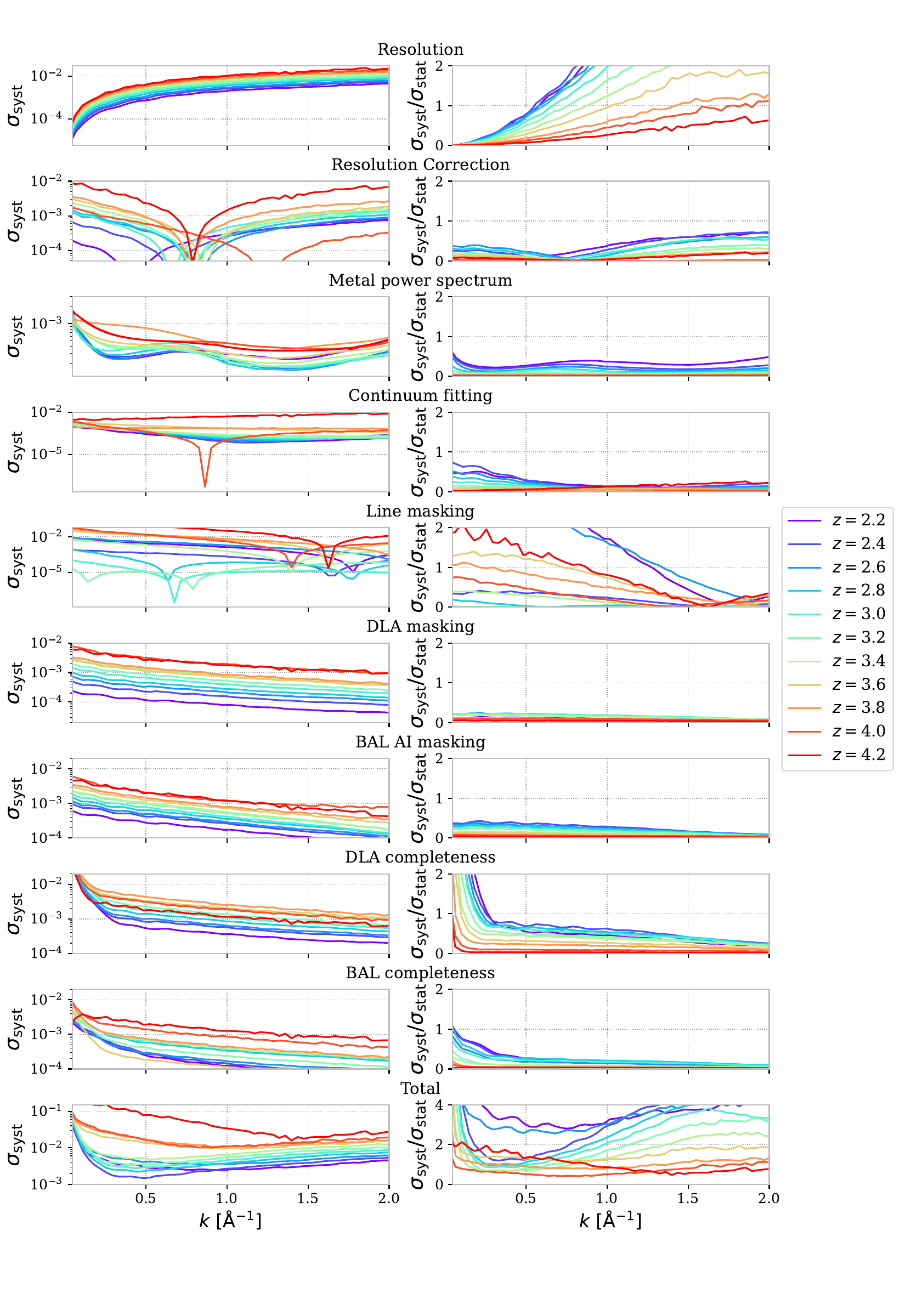}
    \caption{Systematic uncertainties of the \drone~$\pk$ measurement computed from the different sources considered in this study. The left panel shows the absolute level of those uncertainties, while the right panel shows their level relative to statistical uncertainties (see figure \ref{fig:stat_error}). Each systematic uncertainty is shown in a different line of panels, and the total systematic error, computed in quadrature, is shown on the bottom panels.}
    \label{fig:syst_error}
\end{figure}

We re-evaluated the systematic uncertainty budget of \desiedrfft~to account for new sources of systematics and improvements developed in this article and in the validation companion paper \validpaper. More specifically, pixel masking and resolution corrections were reevaluated to match \drone~statistics. The continuum fitting correction, see below, was also derived in the same paper.

All systematic uncertainties included in the total error budget are shown in figure \ref{fig:syst_error}. As mentioned in section \ref{subsec:noise}, with the development of a cross-exposure estimator that measures the noise in a fully data-driven way, we consider that the systematic error associated with the noise level estimation is negligible and can safely be removed from the systematic error budget. Other systematic uncertainties associated to a bin $(A,z)$ and noted $\sigma_{\mathrm{syst,X}}$, are defined similarly to \desiedrfft~in the following way:

\begin{itemize}
	\item\textbf{{Resolution damping}}:\enspace We ascribe an uncertainty to the stability of the Point Spread Function (PSF) and its impact on the DESI resolution matrix. Following \desiedrfft, the average resolution damping in figure~\ref{fig:resolution_characterization} is fitted with a simplified resolution damping model $\exp \left(-0.5(k \Delta \lambda)^2\right) \cdot \operatorname{sinc}\left(0.5 k \Delta \lambda_{\text {pix }}\right)$ to measure the effective spectral resolution $\Delta \lambda$ in \AA. The $\pk$~uncertainty is given by $\sigma_{\mathrm{syst,res}}(A,z) = 2k_{A}^{2} \Delta \lambda \sigma_{\Delta\lambda} \cdot \pk(A,z)$, where $\sigma_{\Delta\lambda}$ is the error on $\Delta\lambda$ for which we take a conservative absolute error of $1\%$ following the PSF stability measurement in \cite{guy_spectroscopic_2022}.
	\item\textbf{{Resolution correction}}:\enspace A correction for the resolution modeling $A_{\mathrm{res}}(A,z)$ is derived in the companion paper \validpaper. We apply this multiplicative correction to the main $\pk$ measurement and ascribe a systematic error that is $30$ \% of the correction, i.e.,  $\sigma_{\mathrm{syst,res}}(A,z) = 0.3 \left|A_{\mathrm{res}}(A,z) - 1 \right| \pk(A,z)$.
	\item\textbf{{Metal power spectrum}}:\enspace The metal power spectrum is measured in section \ref{subsec:drone_update} in the SB1 side-band wavelength range and shown in figure~\ref{fig:metal_power}. The $\pk$ measurement is directly corrected by a physically motivated fit $\psbmz$. We add a systematic uncertainty equal to the statistical error bar of the side-band power spectrum.
	\item\textbf{{Continuum fitting}}:\enspace A multiplicative correction accounting for the bias introduced by the continuum fitting is derived from \drone-like mocks in \validpaper. This correction, noted $A_{\mathrm{cont}}(A,z)$, directly multiplies the power spectrum measurement, and a $30$ \% relative systematic uncertainty is added to the total budget.
	\item\textbf{{Pixel masking corrections}}:\enspace At the continuum fitting stage, some spectrum pixels are masked due to DLA, BAL passing the $AI > 0$ criterion, and atmospheric emission lines, as discussed in sections \ref{sec:data} and \ref{subsec:dla_bal_split}. Multiplicative corrections, respectively noted $A_{\mathrm{dla}}(z)$, $A_{\mathrm{bal}}(A,z)$, and $A_{\mathrm{line}}(A,z)$ are computed in \validpaper. As for the previous multiplicative corrections, we correct the power spectrum measurement and associate a $30$ \% relative systematic uncertainty. For the special case of BAL masking, a dedicated study in \validpaper~shows that a 30 $\%$ value is over-estimating the systematic uncertainty, and we take a 6 $\%$ value.
	\item\textbf{{Catalog completeness}}:\enspace The DLA and BAL finding algorithms detailed in section \ref{sec:data} result in incomplete detections, especially for the low $\snr$ spectra. The data splits presented in section \ref{subsec:dla_bal_split} provide insight into the impact of undetected DLA and BAL. In contrast with \desiedrfft, we derive the impact of DLA and BAL from the data splits instead of mocks, making this analysis independent of the distribution of DLA and BAL in mocks. For DLA, we compare the power spectrum obtained when DLA are masked and the power spectrum corrected by $A_{\mathrm{dla}}(z)$ to the case where no DLA are masked. The resulting ratio corresponds to the top left panel of figure \ref{fig:dla_bal_splits} with the additional DLA masking correction. We fit this ratio with a power-law function provided by \cite{rogers_simulating_2017}. Following \desiedrfft~which uses the same DLA finding algorithms, we ascribe a conservative $20$ \% incompleteness to the DLA finders. The associated systematic uncertainty is obtained by multiplying this incompleteness factor by the product of the fitted function times $\pk$. In a very similar way, we estimate the impact of unmasked $AI$ BAL absorptions by comparing our new baseline for which we masked $AI$ BAL, corrected by $A_{\mathrm{bal}}(A,z)$, to the previous baseline measurement where those absortions where left unmasked. We use the same fitting function and associate a systematic uncertainty considering a $15$ \% incompleteness for the finder, following \cite{Filbert2023}.
\end{itemize}

For multiplicative correction, the $30$ \% relative systematic uncertainty is motivated by considering a random shift of the correction between $0$ and $100$ \%. Considering a uniform distribution for this shift gives a $30$ \% standard deviation. It is a straightforward but very conservative way of generating an error bar associated with the multiplicative corrections because it gives overestimated error bars related to larger corrections, such as atmospheric lines. In contrast, taking the standard deviation of the correction for different mock realizations gives too small and non-conservative error bars. We chose not to use this method based on mocks for systematic uncertainties.

All systematic uncertainties are added in quadrature to compute the total systematic error, which is shown in the last panels of figure~\ref{fig:syst_error}. The correlations between wavenumber bins induced by these systematics are complex to model, and their contribution to the covariance matrix will be discussed in the upcoming inference of this work. Given the improvement in statistics compared to \desiedrfft, the statistical error bars have decreased by between 260 to 300 \% depending on redshift, and now the $\pk$ measurement is dominated by systematic errors for most redshifts and wavenumbers.

\section{Measurement}
\label{sec:measurement}

\begin{figure}
    \centering
    \includegraphics[width=\linewidth]{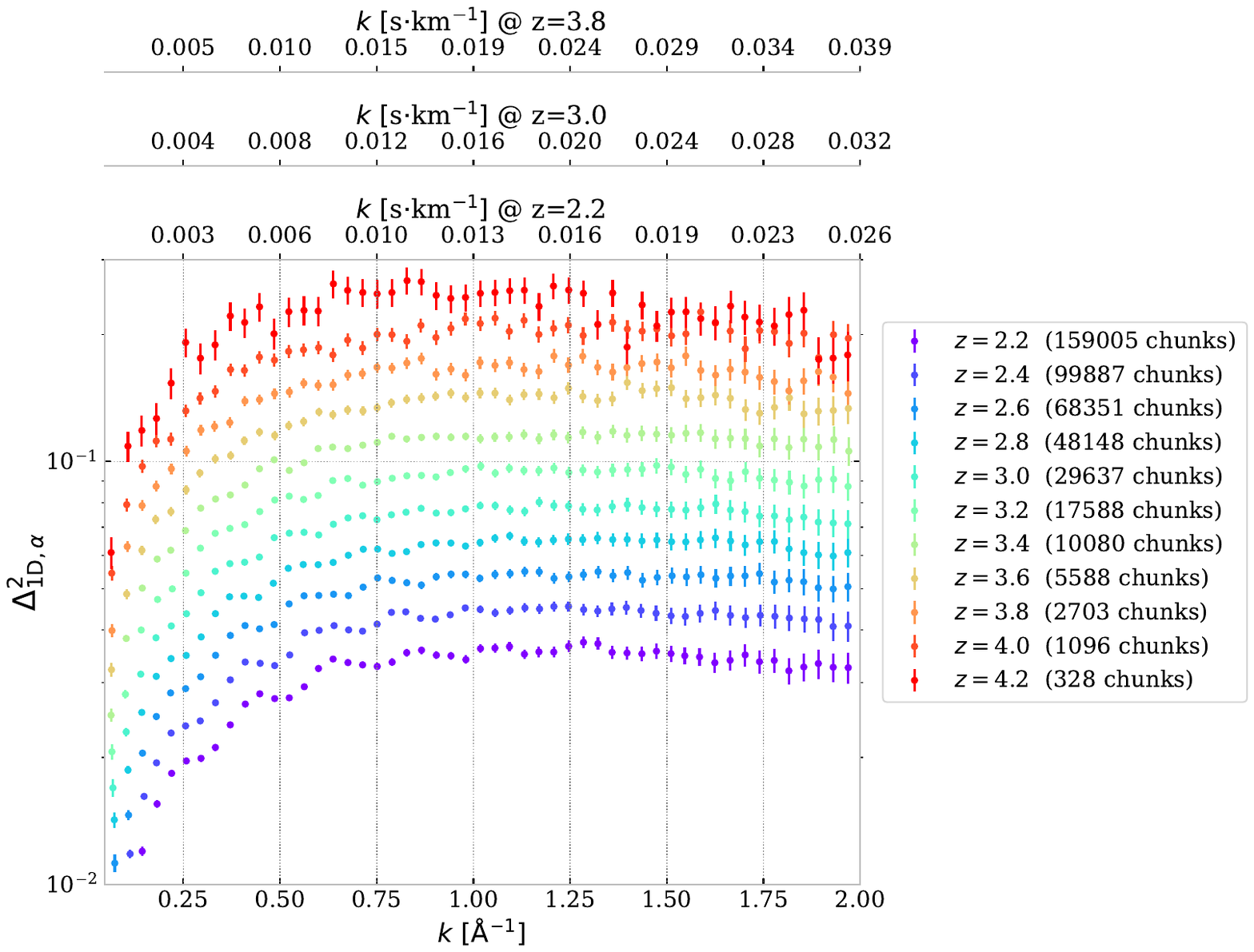}
    \caption{The normalized one-dimensional \lya~forest power spectrum ($\Delta_{1\mathrm{D},\alpha}(k) = k \pk/\pi$) obtained from the \dronefull~data set using the FFT $\pk$ estimator given in figure~\ref{eq:final_p1d_estimator}. This power spectrum is computed from the new baseline which includes BAL $AI$ absorption masking. For homogeneity reasons, we show the measurement without the cross-exposure estimator for low redshifts, even if we consider it in the new baseline (see figure~\ref{fig:cross_exposure} for more details). The error bars reported are obtained by adding in quadrature the statistical and systematic uncertainties shown in figures~\ref{fig:stat_error} and~\ref{fig:syst_error}. The measurement is shown with \AA~units but the equivalent $\kms$ units are shown for different redshift on the top of the figure. The number of sub-forest used (noted here chunks) are reported for each redshift bin.}
    \label{fig:p1d}
\end{figure}

\begin{figure}
    \centering
    \includegraphics[width=\linewidth]{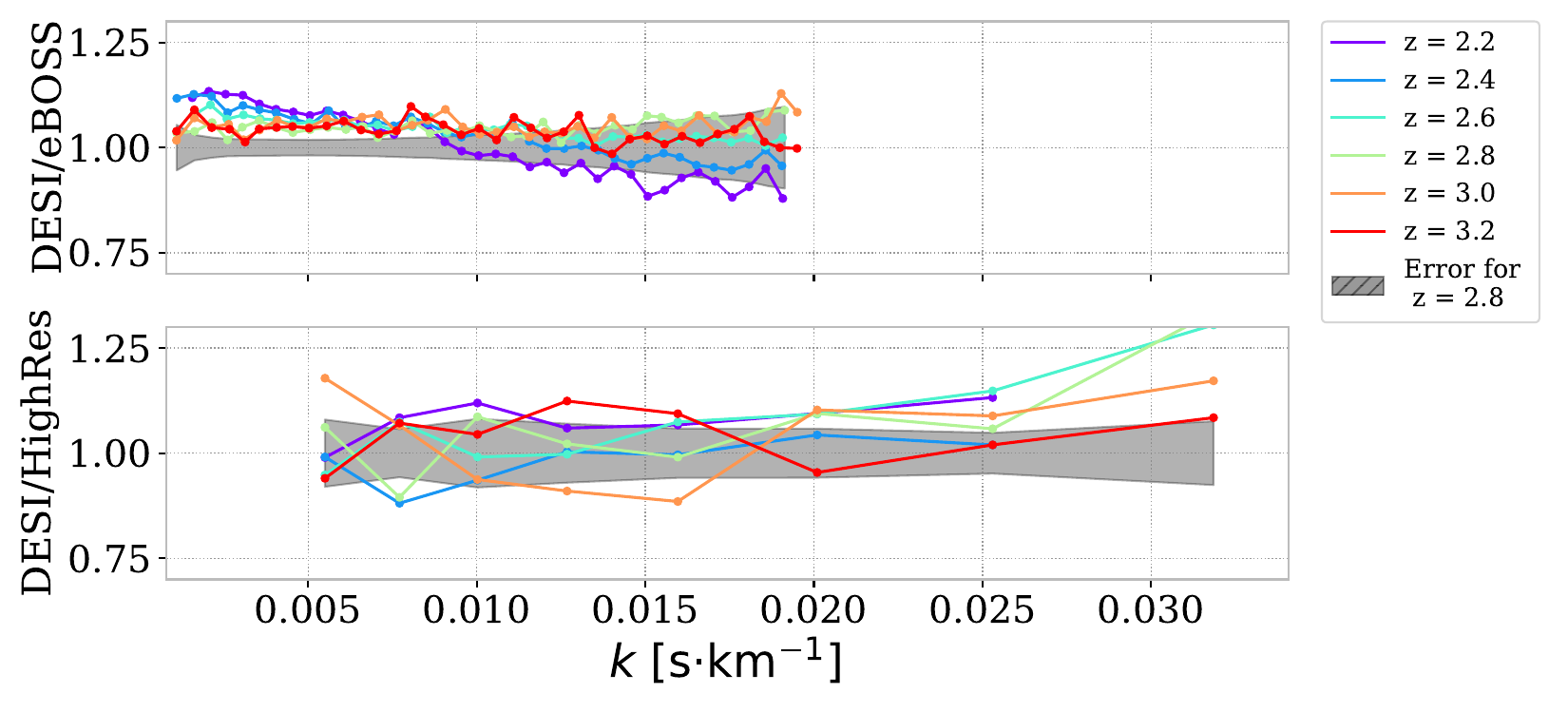}
    \caption{Ratios of our \drone~one-dimensional power spectrum measurement to original measurements from eBOSS data~\cite{chabanier_one-dimensional_2019} (top) and to the combination of KODIAQ, SQUAD, and XQ-100 high-resolution surveys~\cite{karacayli_optimal_2022} (bottom). For clarity, only the ratio error bar associated to the $z=2.8$ redshift bin is shown with the shaded area. This error bar contains the combination of our statistical and systematic uncertainties with the compared $\pk$ error bars. For the high-resolution comparison, our $\pk$ measurement and the error bars are rebinned to the same wavenumber binning.}
    \label{fig:p1d_comparison}
\end{figure}

The $\pk$ measurement is obtained from running the entire pipeline given in \ref{sec:method}, correcting for underestimated noise for high redshifts ($z > 2.7$), using the cross-exposure estimator for low redshift ($z < 2.7$), applying pixel masking, continuum fitting and resolution corrections, and subtracting the estimated metal power spectrum $\psbmz$. The final $\pk$ estimator is defined as

\begin{equation}
\label{eq:final_p1d_estimator}
\begin{aligned}
\pk(A,z) =& A_{\mathrm{line}}(A,z) A_{\mathrm{dla}}(z) A_{\mathrm{bal}}(A,z) A_{\mathrm{cont}}(A,z) A_{\mathrm{res}}(A,z) \\
& \times \left(\left\langle \ps(k) \right\rangle_{s \in z, k\in A} - \psbm(A,z) \right)\,,
\end{aligned}
\end{equation}

\noindent where the individual power spectra definition is redshift dependent:

\begin{equation}
	\ps (k)= \left\{\begin{array}{ll}
\mathcal{P}_X\left[ \overrightarrow{\frac{\dfs(k)}{\rmats(k)}}, \overrightarrow{\frac{\dfs(k)}{\rmats(k)}} \right]  & \text{for } z < 2.7, \\
\left[\prs(k)-\pns(k) - \alpha \right] \cdot \rmats^{-2}(k) & \text{for } z > 2.7 \, . \end{array}\right. 
\end{equation}

Figure~\ref{fig:p1d} shows the one-dimensional power spectrum measurement obtained from \dronefull~data set with this full FFT estimator. For homogeneity reasons, the cross-exposure is not shown on the figure; we refer the reader to figure~\ref{fig:cross_exposure} for the difference between the two estimators. The error bars are the quadratic sum of the statistical and total systematic uncertainties detailed in previous sections. The increase in statistics is visible by eye compared to the previous \edrmtwo~measurement in \desiedrfft. We did not compute the $\pk$ for redshift higher than $4.2$ because of the too-low number of sub-forests and the significant variation of $\pk$.

We compare the resulting $\pk$ measurement with the previous eBOSS~\cite{chabanier_one-dimensional_2019} and high-resolution measurements from the combination of KODIAQ, SQUAD, and XQ-100 surveys~\cite{karacayli_optimal_2022} in figure~\ref{fig:p1d_comparison}. As those studies expressed $\pk$ in velocity units ($\invkms$), the \drone~power spectrum is converted using $k \left[\kms\right] = k\left[\text{\AA}^{-1}\right] \times \lambda_{\alpha} (1+z) / c$ before the $\pk$ averaging stage using the redshift of individual sub-forests. We reproduce all aforementioned corrections with the velocity unit measurements on \drone~data and synthetic data in \validpaper.

\begin{figure}
    \centering
    \includegraphics[width=\linewidth]{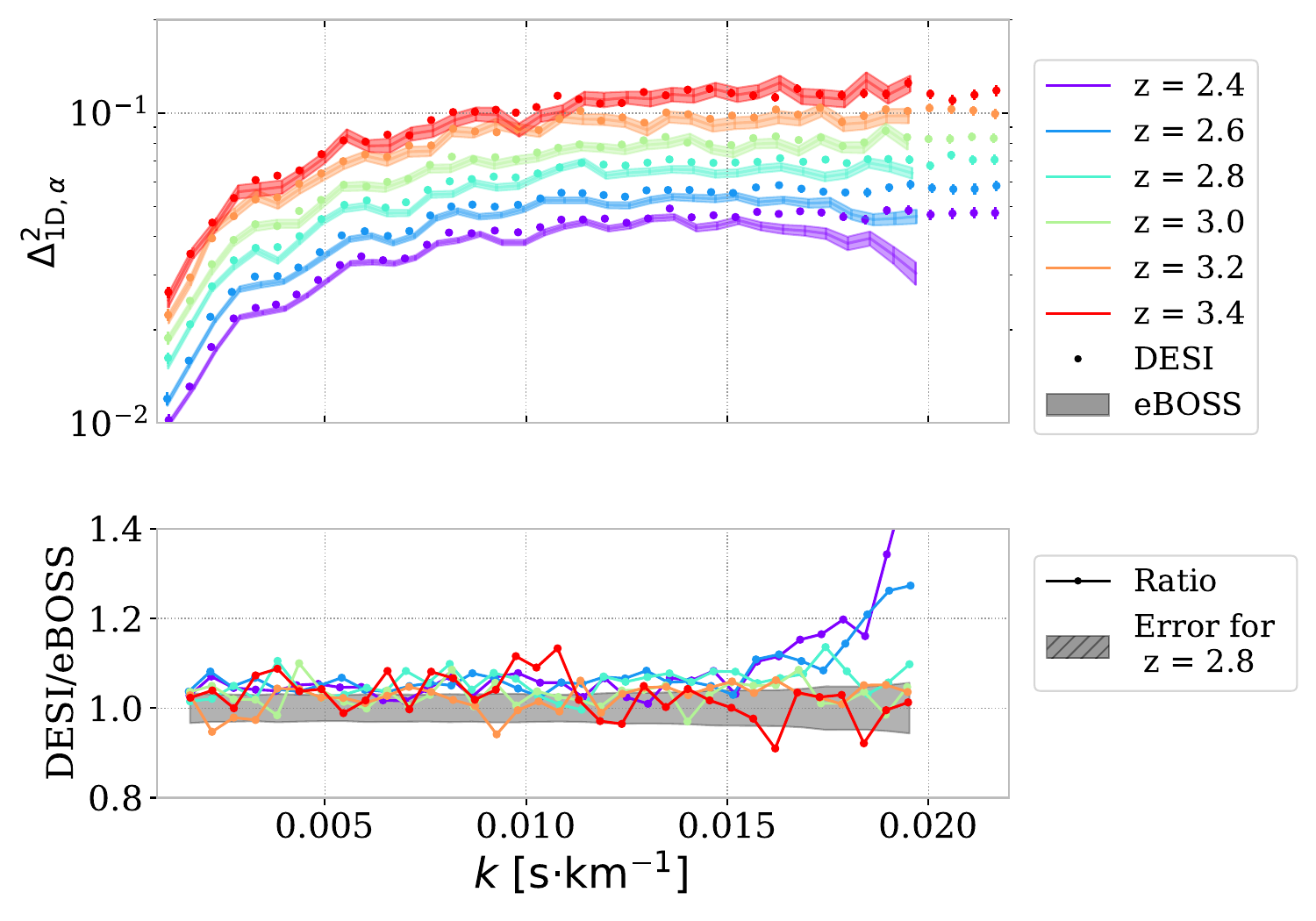}
    \caption{(top) eBOSS (shaded area) and DESI \drone\ (points) $\pk$ measurements resulting from a similar analysis pipeline and using the DESI DLA catalog. (bottom) Ratio of DESI \drone~over this updated eBOSS measurement.}
    \label{fig:eboss_desi}
\end{figure}

We observe a discrepancy with eBOSS, especially at large scales ($k < 0.005\ \invkms$) and for the first two redshift bins ($z = 2.2, 2.4$). This is primarily due to the differences in the DLA catalogs. Indeed, analyzing eBOSS data with our DLA catalog and using procedures close to the DESI analysis results in a DESI to eBOSS ratio independent of $z$ and $k$ for $k<0.015$ s/km, as illustrated in figure~\ref{fig:eboss_desi}. 
The fact that the ratio is independent of $k$ does not indicate an issue in the treatment of resolution or noise. We investigated the possibility that the discrepancy arises from the unit conversion between velocity (eBOSS) and \AA~(DESI), but could not find any issue there. We also considered the possibility that a sky residual in eBOSS spectra reduces the flux contrasts and then the power spectra. In this case, the effect would be more important for low amplitude spectra, but we did not observe any dependence of the DESI to eBOSS ratio on the spectra amplitude. Consequently, the origin of this 1.05 ratio is so far unknown and will be further studied in future publications. Since this ratio is quite wavenumber independent, we anticipate that it will mainly impact power spectrum amplitude parameters such as $\sigma_8$. It should have a minor impact on wavenumber-dependent parameters such as matter power spectrum slope, neutrino masses and warm dark matter constraints.
We ascribe the differences at higher $k$ in figure~\ref{fig:eboss_desi} to the noise characterization, which is much less controlled for eBOSS than for DESI data set. This is confirmed by the fact that DESI data appear in agreement with high-resolution experiments up to $k<0.02$ s/km in figure \ref{fig:p1d_comparison}.

The difference between our \drone~measurement and previous non-DESI ones is smaller than for \edrmtwo~measurement in~\desiedrfft. In particular, our measurement is in better agreement with the high-resolution measurement at the smallest scales, implying an improved control of small-scale systematics in our study. However, we note that there still exists a disagreement larger than the error bar at $k > 0.02\ \kms$ which will need to be explored in future work before using both measurements for a cosmological interpretation. Since the DESI spectroscopic pipeline is built on a more robust methodology (e.g., "spectroperfectionism" algorithm detailed in \cite{bolton_spectro-perfectionism_2010}) than the eBOSS one, and considering the extended data split analysis performed in our study, we are confident that our \drone~measurement is more robust than eBOSS for cosmological interpretation.

\section{Conclusion and prospects}
\label{sec:conclusion}

Using a Fast Fourier Transform estimator, we measured the one-dimensional \lya~power spectrum with the \dronefull~data set. Following the work performed in the previous DESI study \desiedrfft~\cite{Ravoux2023}, we extended the instrumental characterization to the entire \drone~data set. The noise control was improved by introducing a new $\pk$ estimator, which uses the individual cross-power spectrum between different DESI exposures of the same quasar. This cross-exposure estimator, used for the low redshift sample of the \drone~measurement, does not require explicit modeling of the noise power spectrum and allows in addition for the reduction of fiber-related instrumental effects.

An extended analysis of potential sources of systematics was performed by computing $\pk$ for various data splits and method variations. Those studies notably concluded that the broad absorption line quasars detected with the absorption index should be accounted for by masking the associated absorptions. All the other data splits and method variations highlighted the robustness of our measurement. We conducted a detailed investigation of the statistical and systematic uncertainty budget. Notably, the variance estimator had to be modified to take into account spectrum weighting, and the impact of broad absorption line quasars was added to the systematic budget. A covariance estimator that accounts for $\snr$ weighting was developed and used to compute the covariance matrix, which will be used for cosmological interpretation of the measurement.

The resulting $\pk$~measured on the \drone~data set constitutes the best intermediate-resolution power spectrum measurement in terms of resolution and statistics. In particular, our measurement represent a drastic statistical improvement ($5$ times increase in number of quasars, and around $2.8$ times decrease in statistical uncertainty) compared to the previous DESI measurement~\cite{Ravoux2023,karacayli_optimal_2023}. We compared the \drone~measurement to eBOSS~\cite{chabanier_one-dimensional_2019} and high-resolution~\cite{karacayli_optimal_2022} measurements. Relative to \desiedrfft, we observe a global improvement in the agreement with previous measurements, although we still have large and small-scale discrepancies. 

The FFT measurement is associated with a companion paper~\cite{karacayli2025} presenting the measurement of $\pk$ with a quadratic maximum likelihood estimator. Both measurements agree in a comparison made in~\cite{karacayli2025}. They will be cosmologically interpreted in another paper in preparation. High-resolution hydrodynamic simulations will be used in hand with state-of-the-art Gaussian process emulators~\cite{pedersen_emulator_2020,walther_simulating_2021,Walther2024} to interpret those $\pk$ measurements. They will put strong constraints on the amplitude and slope of the matter power spectrum, the sum of neutrino masses, and exotic dark matter models. 

The uncertainty budgeting conducted in this article showed that systematic uncertainties now dominate the one-dimensional power spectrum measurement for most scales. Future measurements should aim to enhance the characterization of the different systematic sources to lower their contribution, as done for the noise systematic in this study. The increase in statistics with the next data release of DESI will provide the opportunity to measure $\pk$ data from sub-samples less contaminated by systematics. For example, the systematics related to pixel masking can be suppressed using masking window methods as in~\cite{Lokken2025}. We note that other surveys such as 4MOST Cosmology Redshift Survey~\cite{Richard2019} or WEAVE-QSO~\cite{pieri_weave-qso_2016} will also provide clean \lya~forest data sets that can be used to measure $\pk$.

\appendix

\section{Updated figures for \drone~data set}
\label{appendix:updated_plots}

\begin{figure}
    \centering
    \includegraphics[width=0.8\linewidth]{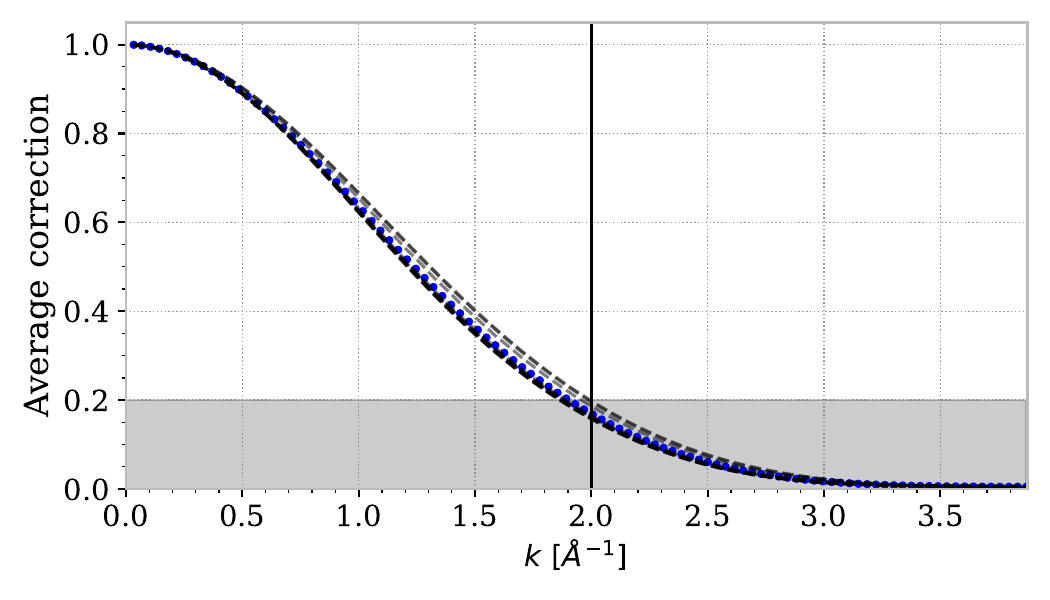}
    \caption{Resolution damping ($\left\langle \rmats^{2}(k) \right\rangle_{s \in z, k\in A}$) computed for the \drone~data set. This damping is shown in dashed black lines for the different redshifts and is very weakly dependent on it. Blue points represent the average over all redshifts. The shaded region represents the wavenumbers for which the impact due to resolution damping is larger than $80$ \%. Using this criterion, the vertical line defines the maximal wavenumber used in our \drone~analysis.}
    \label{fig:resolution_characterization}
\end{figure}

\begin{figure}
    \centering
    \includegraphics[width=\linewidth]{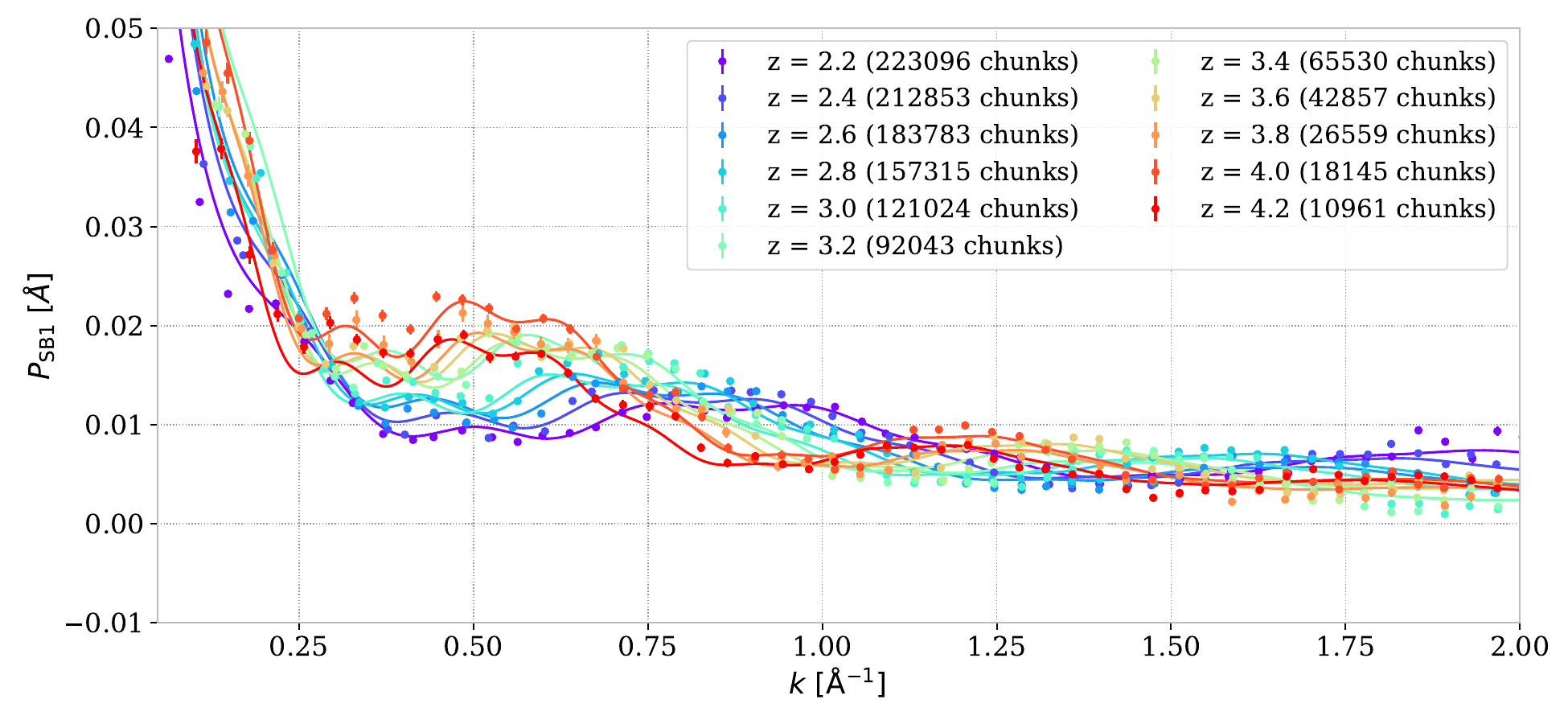}
    \caption{One-dimensional power spectra measured in the side-band region SB1 of \drone~data set with noise correction adapted to SB1 region. The noise power spectrum used for this SB1 calculation is corrected with an additive term $\alpha = 0.00018$ \AA, measured with the same method performed in figure~\ref{fig:noise_power} adapted to SB1. The continuous lines are obtained by fitting the model in equation \ref{eq:side_band_model} with a first-order redshift-dependent polynomial function to the data points represented by points. This fitted model is used to correct for the metal power spectrum in the main measurement.}
    \label{fig:metal_power}
\end{figure}

This appendix contains the figures for which we used in this paper the same methods as were used for \edrmtwo~data set in \desiedrfft. 
We provide those figures to show an update using the new \dronefull~data set. Figure~\ref{fig:resolution_characterization} shows the resolution damping computed as $\left\langle \rmats^{2}(k) \right\rangle_{s \in z, k\in A}$ from the average of the Fourier transform of the DESI pipeline resolution matrix. This resolution damping determines both the maximal wavenumber considered and the systematic uncertainty from PSF stability.

Figure~\ref{fig:metal_power} shows the power spectrum measured in the side-band SB1 ($1,270 < \lambda_{\mathrm{rf}} < 1,380$ \AA) and used to estimate the metal power spectrum $\pme$.

\section{Covariance and variance estimators}
\label{appendix:cov}

Let us consider we have $n$ pairs of measurements $(X_i,Y_i)$ of two physics quantities $X$ and $Y$ and the weighted means $\overline X =  \sum_{i=1}^n v_i X_i / \sum_{i=1}^n v_i$ and  $\overline Y =  \sum_{i=1}^n w_i Y_i / \sum_{i=1}^n w_i$.
We note $E(X_i)=\mu$, $E(Y_i)=\nu$ and $\cov(X_i,Y_j)=\delta_{ij} c_i$, which means we assume $\cov(X_i,Y_j)=0$ when $i\ne j$ and in particular $\cov(X_i,X_j)=0$. We have

\begin{equation}
\label{eq:truecov}
\cov(\overline X, \overline Y) =  \frac{\sum_{ij} v_i w_j \cov(X_i,Y_j)}{\sum_i v_i \sum_i w_i} = \frac{\sum_i v_i w_i c_i}{\sum_i v_i \sum_i w_i} \, .
\end{equation}

We want to extract this covariance from the dispersions of the $X_i$ and $Y_i$. In order to get the $\sum_i v_i w_i c_i$ term, we should include $\sum_i v_i w_i X_i Y_i$ in the estimator and normalize it by $\sum_i v_i w_i$. But this will produce a term in $\mu\nu$ that we should try to cancel out, let's try the following quantity:

\begin{equation}
\label{eq:estimcov}
T =    \frac{\sum_i v_i w_i X_i Y_i}{\sum_i v_i w_i} -  \frac{\sum_i v_i X_i }{ \sum_i v_i} \frac{\sum_i w_i Y_i }{ \sum_i w_i}  \, .
\end{equation}

We have

\begin{equation}
E \left( \sum_i v_i w_i X_i Y_i \right) =  \sum_i v_i w_i [\cov(X_i Y_i)+E(X_i)E(Y_i)] =\sum_i  v_i w_i (c_i +\mu\nu) \, ,
\end{equation}

\noindent and

\begin{eqnarray}
E \left( \sum_i v_i X_i  \sum_j w_jX_j \right) & = &  \sum_{ij}  v_i w_i  E(X_i Y_j) =  \sum_{ij} v_i w_j (\mu\nu + \delta_{ij} c_i)  \\
& = & \mu\nu\sum_i w_i \sum_j w_j  + \sum_i v_i w_i c_i  \, .
\label{eq:esp_term2}
\end{eqnarray}

We can now compute $E(T)$ where the $\mu\nu$ terms do cancel out and get

\begin{align}
E(T) =  \frac{\sum_i v_i w_i c_i}{\sum_i v_i w_i}  -  \frac{\sum_i v_i w_i c_i}{\sum_i v_i \sum_j w_i} =   \left( \frac{\sum v_i \sum w_i}{\sum v_i w_i} -1\right)  \frac{\sum_i v_i w_i c_i}{\sum_i v_i \sum_j w_i}  \, .
\end{align}

So we just have to divide $T$ by the above parenthesis to get an unbiased estimator of the covariance:

\begin{equation}
\label{eq:estimcov2}
\widehat \cov =   \left( \frac{\sum v_i \sum w_i}{\sum v_i w_i} -1\right)^{-1}  \left[ \frac{\sum v_i w_i X_i Y_i}{\sum v_i w_i} -  \frac{\sum_i v_i X_i }{ \sum_i v_i} \frac{\sum_i w_i Y_i }{ \sum_i w_i} \right]  \, ,
\end{equation}

\noindent which gives for the variance:

\begin{equation}
\label{eq:estim2var}
\widehat \var =   \left( \frac{(\sum w_i)^2}{\sum w_i^2} -1\right)^{-1}  \left[ \frac{\sum w_i^2 X_i^2 }{\sum w_i^2} -  \left(\frac{\sum_i w_i X_i }{ \sum_i w_i}\right)^2 \right] \, .
\end{equation}

We can set $X_i=P_{A,s}$, $Y_i=P_{B,s}$, $v_i=w_{A,s}$ and $w_i=w_{B,s}$ to get our estimators of the variance and covariance of $\pk$.

An alternative estimator of the variance of a weighted average is sometimes proposed~\cite{Bevington2003}, where $w^2$ is replaced by $w$. However, the same kind of algebra as above shows that this estimator fails if the weights are not proportional to the inverse variance. In our case, this proportionality is only approximate, which makes this estimator inappropriate. In addition, this $w$ estimator cannot be generalized to a covariance estimator. These analytic findings were confirmed by a small Monte Carlo simulation that showed that our $w^2$ estimator for the variance is unbiased and that the $w$ estimator fails in the general case.

\acknowledgments

This material is based upon work supported by the U.S. Department of Energy (DOE), Office of Science, Office of High-Energy Physics, under Contract No. DE–AC02–05CH11231, and by the National Energy Research Scientific Computing Center, a DOE Office of Science User Facility under the same contract. Additional support for DESI was provided by the U.S. National Science Foundation (NSF), Division of Astronomical Sciences under Contract No. AST-0950945 to the NSF’s National Optical-Infrared Astronomy Research Laboratory; the Science and Technology Facilities Council of the United Kingdom; the Gordon and Betty Moore Foundation; the Heising-Simons Foundation; the French Alternative Energies and Atomic Energy Commission (CEA); the National Council of Humanities, Science and Technology of Mexico (CONAHCYT); the Ministry of Science, Innovation and Universities of Spain (MICIU/AEI/10.13039/501100011033), and by the DESI Member Institutions: \url{https://www.desi.lbl.gov/collaborating-institutions}. Any opinions, findings, and conclusions or recommendations expressed in this material are those of the author(s) and do not necessarily reflect the views of the U. S. National Science Foundation, the U. S. Department of Energy, or any of the listed funding agencies.

The authors are honored to be permitted to conduct scientific research on I'oligam Du'ag (Kitt Peak), a mountain with particular significance to the Tohono O’odham Nation.

The project leading to this publication has received funding from Excellence Initiative of Aix-Marseille University - A*MIDEX, a French ``Investissements d'Avenir'' program (AMX-20-CE-02 - DARKUNI). The authors acknowledge support from ANR grant ANR-22-CE92-0037.

\paragraph{Data availability} 
\ 

The DESI spectra and associated catalogs from \dronefull~are publicly available\footnote{\label{dr1data}\url{https://data.desi.lbl.gov/doc/releases/dr1/}}. All the plots of this article are generated with \texttt{picca}$^{\ref{igmhub/picca}}$ (9.17.0) and \texttt{p1desi}~\git{corentinravoux/p1desi}{} (2.0.2). All the data points of the figures in this article will be made available upon publication according to the data management policy of DESI. All the data points of the figures and measurements in this article are made available according to the data management policy of DESI\footnote{\label{zenodo_data_release}\url{https://doi.org/10.5281/zenodo.17100308}}.

\bibliographystyle{JHEP}
\bibliography{biblio}

\end{document}